\documentclass[10pt,journal,compsoc]{IEEEtran}
\usepackage{framed}
\usepackage{listings}
\usepackage{xcolor}
\usepackage{graphicx}
\usepackage{caption}
\usepackage{amsmath}
\usepackage{amsfonts} 
\usepackage{algorithm}
\usepackage{amssymb}
\usepackage{algorithmic}
\usepackage[top=2cm, bottom=2cm, left=2cm, right=2cm]{geometry}
\usepackage{hyperref}
\usepackage{environ}
\usepackage{tikz}
\usepackage{ragged2e}
\usepackage{hyperref}
\usepackage{booktabs}
\usepackage{multicol}
\usepackage{multirow}
\usepackage{threeparttable}
\usepackage{comment}
\usepackage{colortbl}
\usepackage{textcomp}
\usepackage{bbding}
\usepackage{CJKutf8}
\definecolor{c1}{HTML}{7e0f12}

\ifCLASSOPTIONcompsoc
\else
\fi

\ifCLASSINFOpdf
\else
\fi

\begin{document}

\title{AccessFixer: Enhancing GUI Accessibility for Low Vision Users with R-GCN Model}

\author{Mengxi~Zhang,
        Huaxiao~Liu,
        Chunyang~Chen,~\IEEEmembership{Member,~IEEE},
        Guangyong~Gao,
        Han~Li,
        and Jian~Zhao
\IEEEcompsocitemizethanks{\IEEEcompsocthanksitem Mengxi Zhang, Huaxiao Liu (Corresponding author), Guangyong Gao, and Han Li are with the College of Computer Science and Technology, Jilin University, Changchun, China.\protect\\
E-mail: liuhuaxiao@jlu.edu.cn
\IEEEcompsocthanksitem Chunyang Chen is with the Faculty of Information Technology, Monash University, Australia. E-mail: chunyang.chen@monash.edu
\IEEEcompsocthanksitem Jian Zhao is with the Department of Computer Science and Technology, Changchun University, Changchun, China. E-mail: zhaojian@ccu.edu.cn}
\thanks{}}

\markboth{IEEE Transactions on Software Engineering}%
{Zhang \MakeLowercase{\textit{et al.}}: AccessFixer: Enhancing GUI Accessibility for Low Vision Users with R-GCN Model}

\IEEEtitleabstractindextext{%
\begin{abstract}
\justifying
The Graphical User Interface (GUI) plays a critical role in the interaction between users and mobile applications (apps), aiming at facilitating the operation process.
However, due to the variety of functions and non-standardized design, GUIs might have many accessibility issues, like the size of components being too small or their intervals being narrow.
These issues would hinder the operation of low vision users, preventing them from obtaining information accurately and conveniently.
Although several technologies and methods have been proposed to address these issues, they are typically confined to issue identification, leaving the resolution in the hands of developers.
Moreover, it can be challenging to ensure that the color, size, and interval of the fixed GUIs are appropriately compared to the original ones.
In this work, we propose a novel approach named AccessFixer (\textbf{Access}ibility Issues \textbf{Fix}ing Method), which utilizes the Relational-Graph Convolutional Neural Network (R-GCN) to simultaneously fix three kinds of accessibility issues, including small sizes, narrow intervals, and low color contrast in GUIs.
With AccessFixer, the fixed GUIs would have a consistent color palette, uniform intervals, and adequate size changes achieved through coordinated adjustments to the attributes of related components.
Our experiments demonstrate the effectiveness and usefulness of AccessFixer in fixing GUI accessibility issues.
After fixing 30 real-world apps, our approach solves an average of 81.2\% of their accessibility issues.
Compared with the baseline tool that can only fix size-related issues, AccessFixer not only fixes both the interval and color contrast of components, but also ensures that no new issues arise in the fixed results.
Also, we apply AccessFixer to 10 open-source apps by submitting the fixed results with pull requests (PRs) on GitHub. 
The results demonstrate that developers approve of our submitted fixed GUIs, with 8 PRs being merged or under fixing.
A user study examines that low vision users host a positive attitude toward the GUIs fixed by our method.
\end{abstract}

}

\maketitle

\section{Introduction}

\IEEEPARstart{M}{o}bile applications (apps) have become an integral part of people's daily lives, and they are constantly being updated and improved to satisfy the needs of swift information development and expansion.
According to statistics, Google Play has launched more than 2.67 million Android apps~\cite{GooglePlay} as of 2022.
With such numerous apps, people's living habits have changed significantly, as reflected in online shopping, online learning, virtual work, etc.
Such online work, studies, or entertainment could bring significant convenience to sighted people.
However, many visually impaired people have difficulty enjoying these conveniences, especially those with low vision.

Low vision is the eyesight loss that cannot be corrected with glasses, contacts, or surgery, and it is not blindness as limited sight remains~\cite{NationalEye}, as the National Eye Institute introduced~\cite{NationalEyeInstitute}. 
This disease includes blurry sight, blind spots, and poor vision, and it is common in our society~\cite{Xu2020PrevalenceAC}.
As announced by the World Health Organization (WHO) in 2022~\cite{WorldHealthO}, there are 2.7 billion visually impaired people worldwide, with more than 70\% having low vision.

\begin{figure}
\centering
\includegraphics[width=8cm]{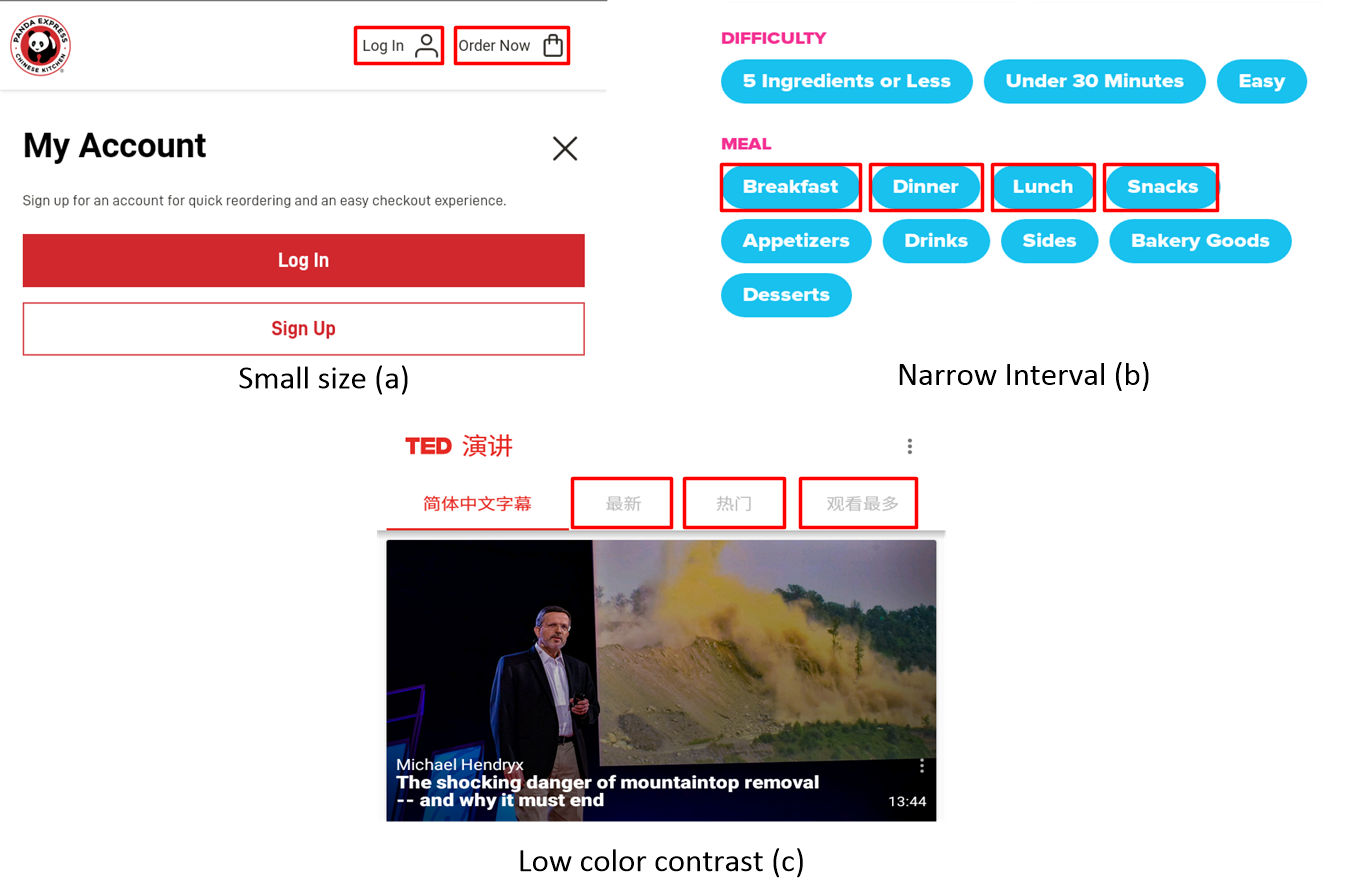}
\caption{The examples of accessibility issues in GUIs for low vision users.}
\label{fig: issues}
\end{figure}
Concerning these users, they mainly rely on their weak eyesight to operate the Graphical User Interface (GUI)~\cite{GUI}, and would inevitably encounter many issues.
Referring to prior studies~\cite{Alshayban2020AccessibilityII}~\cite{Vendome2019CanEU} and the Web Content Accessibility Guidelines (WCAG)~\cite{WCAG}, the components with small size (in Figure~\ref{fig: issues} (a)), narrow intervals (in Figure~\ref{fig: issues} (b)), and low color contrast (in Figure~\ref{fig: issues} (c)) are commonly encountered by these users.
Consequently, such accessibility issues would significantly limit the low vision users in the following two major aspects.
First, these issues would lead users to misoperate with the apps, resulting in the acquisition of wrong messages, information leakage, and other troubles.
Second, users might not identify the GUIs clearly, leading to the omission of crucial information or functionalities.

In the face of such accessibility issues, both industry and academia have proposed various guidelines and methods.
Regarding the industry, Google~\cite{Googleaccessibility} and Apple~\cite{appleaccessibility} issued developer guidelines related to accessibility from the perspectives of apps' reach, versatility, and user needs.
Further, Google released the Accessibility Test Framework~\cite{AccessibilityTest} to explain how the apps would be presented to accessibility services in 2016, and meanwhile, the corresponding tool named Accessibility Scanner (AS)~\cite{AccessibilityScanner} was proposed. 
This tool can scan the GUIs and detect their issues, based on content labels, touch target sizes, clickable items, as well as text and image contrast.
Also, IBM AbilityLab developed the Mobile Accessibility Checker (MAC)~\cite{MobileAccessibilityChecker} that could perform similar functions to AS, but it is weaker than AS in image contrast detection.
As for the academic field, the Programmable UI-automation (PUMA)~\cite{Hao2014PUMAPU}, Mobile Accessibility Test (MATE)~\cite{Eler2018AutomatedAT}, and Xbot~\cite{Chen2021AccessibleON} were proposed to check the problematic components in GUIs.
Among them, PUMA and MATE detected issues in GUIs, such as missing labels, improper sizes, and misplacements, by employing explicit rules followed by the WCAG, but they all lack a focus on color-related issues.
Xbot optimized the traditional tool named Google Monkey~\footnote{\url{https://developer.android.com/studio/test/other-testing-tools/monkey}}, enhancing its capability to detect a wider range of accessibility issues.
The release of these tools has indeed raised developers' awareness of accessibility issues within GUIs to a certain extent. 
However, the results provided by the aforementioned tools typically consist of detection reports indicating problematic components in the GUIs, without offering specific repair strategies. 
The lack of solutions makes it challenging for developers to implement concrete fixes.

Further, with these methods in place, we conduct a preliminary investigation to explore the current status of app accessibility (in Section~\ref{sec: background}).
Unfortunately, among the 500 apps (2,646 GUIs) we investigated, 78.95\% of them have accessibility issues that low vision users might encounter.
Such results indicate that developers still lack attention to the accessibility of GUIs, which are also discussed and confirmed in works of ~\cite{Alshayban2020AccessibilityII}~\cite{Chen2021AccessibleON}~\cite{Vendome2019CanEU}.
Also, developers might confront the following challenges when fixing accessibility issues.
First, they might not conduct requirement research for low vision users, thus, it is unknown how they fix such issues that could satisfy the requirements of these users.
Second, it is hard to maintain a consistent color palette, uniform intervals, and adequate size changes when fixing GUIs to match the original design.
We could imagine that if we only fix the accessibility issues by simply enlarging the size, widening the interval, etc., then the components in the GUIs would present an inconsistent color palette, narrow intervals, and inappropriate sizes, hurting user experience and perception.
To the best of our knowledge, Aalotaib et al.~\cite{Alotaibi2021AutomatedRO} first attempted to fix the size-based issues in GUIs, and kept the visual consistency as much as possible by multi-objective optimization strategy.
However, there are two restrictions in their method after we apply it to real-world GUIs.
One is that it only focuses on size-based issues, but cannot modify the intervals between components, along with the colors of these components.
The other is that their method needs to continuously adjust the algorithm weights to achieve the relative optimality of the four objectives, `Accessibility Heuristic', `Relative Positioning and Alignment of Views', `Minimum Spacing Between Views', and `Amount of View Size Change', to ensure that the repaired component keeps the visual consistency with other components in size.
Still, due to significant variations in the layout of apps across different domains, it can be difficult to objectively measure all apps by a certain weight, which may bring new accessibility issues in the fixed results.

Leveraging this insight, in this work, we propose a novel method, called AccessFixer (\textbf{Access}ibility Issues \textbf{Fix}ing Method).
It breaks the limitation of existing GUI accessibility detection tools that merely provide issue reports, by being capable of fixing accessibility issues within GUIs.
First, we design a strategy that can convert the containers, visible components, and their relations in the GUIs into GUI-graphs.
Second, we construct a pre-trained model based on the Relational-Graph Convolutional Neural Network (R-GCN), and adopt the well-accessible GUIs annotated by low vision users or existing detection tools to train this model, which can better reflect the needs of low vision users and accessibility standards.
Finally, we capture the feature representation of node information within graph structures from the trained model, and establish the mapping set between such representations and the size, interval, and color contrast of components.
This mapping set records and preserves a one-to-one correspondence between the feature representation of nodes and their attributes.
We also provide an approach to adjust the various attributes of the components in the GUIs to be fixed.
In this process, our method can simultaneously consider multiple attribute values of all relevant nodes, which enables it to not only fix the issues in terms of small size, narrow interval, and color contrast, but also ensure their appropriate performance without bringing any extra low vision-related accessibility issues.
More details of this approach are discussed in Section~\ref{sec: ARBot}.
 
We evaluate the effectiveness and usefulness of our method, outlined in Section~\ref{sec: evaluation}.
As for the effectiveness of AccessFixer, we carry out two experiments in Section~\ref{sub: effectiveness}.
One is to utilize existing detection tools (e.g., AS and Xbot) to check the accessibility differences between the fixed GUIs using the AccessFixer and the original GUIs.
The other is to compare our method with the baseline tool proposed by Aalotaib et.al~\cite{Alotaibi2021AutomatedRO} .
The results demonstrate that the number of accessibility issues has dropped by an average of 81\% after re-checking these fixed GUIs.
Further, compared to this baseline tool, AccessFixer could also fix the issues of narrow intervals and low color contrast in GUIs, and would not bring any extra issues, together with fixing 3.54\% more components with size-related issues than the baseline.
To evaluate the usefulness of AccessFixer, in Section~\ref{sub: usefulness}, our experiments involve generating fixed results and submitting them to 10 open-source applications. 
Encouragingly, 8 of them are well processed and have been merged or are under fixing.
After that, we conduct a user study to discuss whether the fixed GUIs could be approved by low vision users.
The results indicate that low vision users possess a highly positive attitude toward the fixed GUIs.
With these experiments, we explore that AccessFixer effectively fixes accessibility issues in GUIs, outperforming the baseline tool, and its usefulness is acknowledged by both developers and low vision users.

To summarize, in this paper, we make the following contributions:
\begin{itemize}
\item A novel method to provide concrete adjustments for GUI components that suffer from accessibility issues in terms of small size, narrow intervals, and low color contrast, while ensuring a consistent color palette, uniform interval, and adequate size.
\item Our method is implemented with the participation of low vision users, and the fixed GUIs are more in line with users' experience.
\item We open-source our datasets of GUIs and codes, which can support subsequent research and exploration.
\end{itemize}

The remainder of this paper is organized as follows:
Section~\ref{sec: related work} explores the related work of our method.
Section~\ref{sec: background} introduces the preparation research we conduct.
Section~\ref{sec: ARBot} presents in detail how the AccessFixer is designed and implemented.
Section~\ref{sec: evaluation} examines the effectiveness of our method, and discusses whether our proposed AccessFixer is useful for developers and users.
Section~\ref{sec: validity} explains the threats to validity of our approach.
Section~\ref{sec: conclusion} draws the conclusion of this work.

\section{Related Work}\label{sec: related work}

\subsection{Research on accessibility in mobile apps}\label{sub: research1}
More recently, there are various related researches on exploring the accessibility in mobile apps~\cite{Alshayban2020AccessibilityII}~\cite{Chen2020UnblindYA}~\cite{Tomlinson2016TalkinAT}.
All of them comply with the WCAG~\cite{WCAG} standard, and aim to guide the development of app accessibility.

Exploring the quantity and types of accessibility issues in apps is a hot concern in the accessibility research community.
Thus, Alshayban et al.~\cite{Alshayban2020AccessibilityII} systematically analyzed the accessibility issues in current apps from the state of affairs, sentiments, and ways forward.
To improve the accessibility of apps, Chen et al.~\cite{Chen2020UnblindYA} proposed a method to generate the labels of icons based on machine learning models.
Further, Mehralian et al.~\cite{Mehralian2021DatadrivenAR} optimized this method by incorporating the contextual information of labels, and obtained more accurate results. 
Except for such methods of generating labels, other studies focused on designing and developing tools that could check accessibility issues. 
To that end, Chen et al.~\cite{Chen2021AccessibleON} developed a tool named Xbot based on large-scale empirical research and the Accessibility Test Framework. 
This tool can capture the accessibility issues of small size and the lack of content-description in GUIs by optimizing the Monkey, and it can also collect more issues than other detection tools. 
Similar tools also include the Mobile Accessibility Test (MATE) by Eler et al.\cite{Eler2018AutomatedAT} and Programmable UI-mobile Automation (PUMA) by Hao et al.~\cite{Hao2014PUMAPU}. 
Both of these methods rely on rule-based approaches to detect accessibility issues within GUIs. 
MATE demonstrates higher accuracy in identifying component size issues due to the diversification of rules specifically tailored for this problem. 
On the other hand, PUMA excels in optimizing various rules, allowing it to identify a broader range of accessibility issues, especially those related to image views. 
Besides, in the industry, Google Research developed Accessibility Scanner~\cite{AccessibilityScanner}.
This tool can scan the GUIs and provide suggestions to improve the accessibility of apps, based on content labels, touch target size, clickable items, as well as text and image contrast.
Also, IBM released Mobile Accessibility Checker~\cite{MobileAccessibilityChecker} to check accessibility issues in GUIs, but it is weak to process the color-related issues.
The biggest advantage of both industry tools is that they can run in real-time on mobile devices and provide reasonable problem reporting.
However, when compared to manual methods, these detection techniques may identify invisible views, while potentially overlooking certain problematic views and identifying invisible ones, such as `RecyclerViews' and `BoxLayouts'.

The above-mentioned studies primarily focus on detecting and reporting the GUI components experiencing issues, such as inconsistent color palettes, narrow intervals, and inappropriate sizing.
Nevertheless, none of them provides solutions to guide developers on tackling these issues.
Confronted with such outcomes, developers are still faced with the confusion of what adjustments should be made for specific components in the repair process.
In many instances, a multitude of accessibility issues can potentially discourage developers, or even abandon their remediation efforts.
As such, a method that involves offering adjustment strategies at a detailed level of component attributes is required for guiding developers in fixing accessibility issues, thereby enhancing the experience for low vision users.
This demand also underscores the necessity for the method proposed in this paper.

\subsection{Research on fixing accessibility issues}\label{sub: research2}
GUIs with unclear layouts and poor visual presentation would significantly impede low vision users from using apps.
Therefore, designing well-accessible GUIs and resolving accessibility issues are crucial means to improve the experience of low vision users.

Currently, there is a lack of research in this area, with only a few studies focused on exploring how to fix accessibility issues or provide practical solutions.
One such study by Li et al.~\cite{Li2020AutoCO} proposed an automatic generation tool for the skeleton layout based on the Transformer model~\cite{Li2020AutoCO}, which solves three issues that developers may introduce in GUI design: improper component positioning, significant variations in component sizes, and disordered hierarchical structures. 
However, this approach can only guide users on how to design new GUIs and cannot be applied to already released GUIs with accessibility issues, nor can it directly repair these problems.
Apart from this tool, the method proposed by Alotaibi et al.~\cite{Alotaibi2021AutomatedRO} in 2021 is the most relevant work to date.
Their method aimed at providing adjustments for size-based accessibility issues in GUIs. 
In more detail, they adopted the multi-objective optimization strategy in the genetic algorithm, and then, optimized four objectives: `Accessibility Heuristic,' `Relative Positioning and Alignment of Views,' `Minimum Spacing Between Views,' and `Amount of View Size Change,' to ensure that the repaired component keeps the visual consistency with other components in size. 
However, due to its exclusive focus on size-related issues, this approach may inadvertently introduce other problems during the fixing process (e.g., reducing the intervals after increasing component sizes). 
More details about this limitation can be found in our experiment of Section~\ref{sub: comparable}.
Such outcomes introduce uncertainty regarding whether the accessibility of the GUI is indeed improved.

The aforementioned methods illustrate the efforts made by researchers in fixing accessibility issues.
However, there is still a need to determine how to effectively fix a diverse array of GUI accessibility issues and to adopt a holistic approach to prevent the emergence of new issues during the repair process.
This is precisely the objective achieved by the approach proposed in this paper.

\begin{figure}
\centering
\includegraphics[width=8cm]{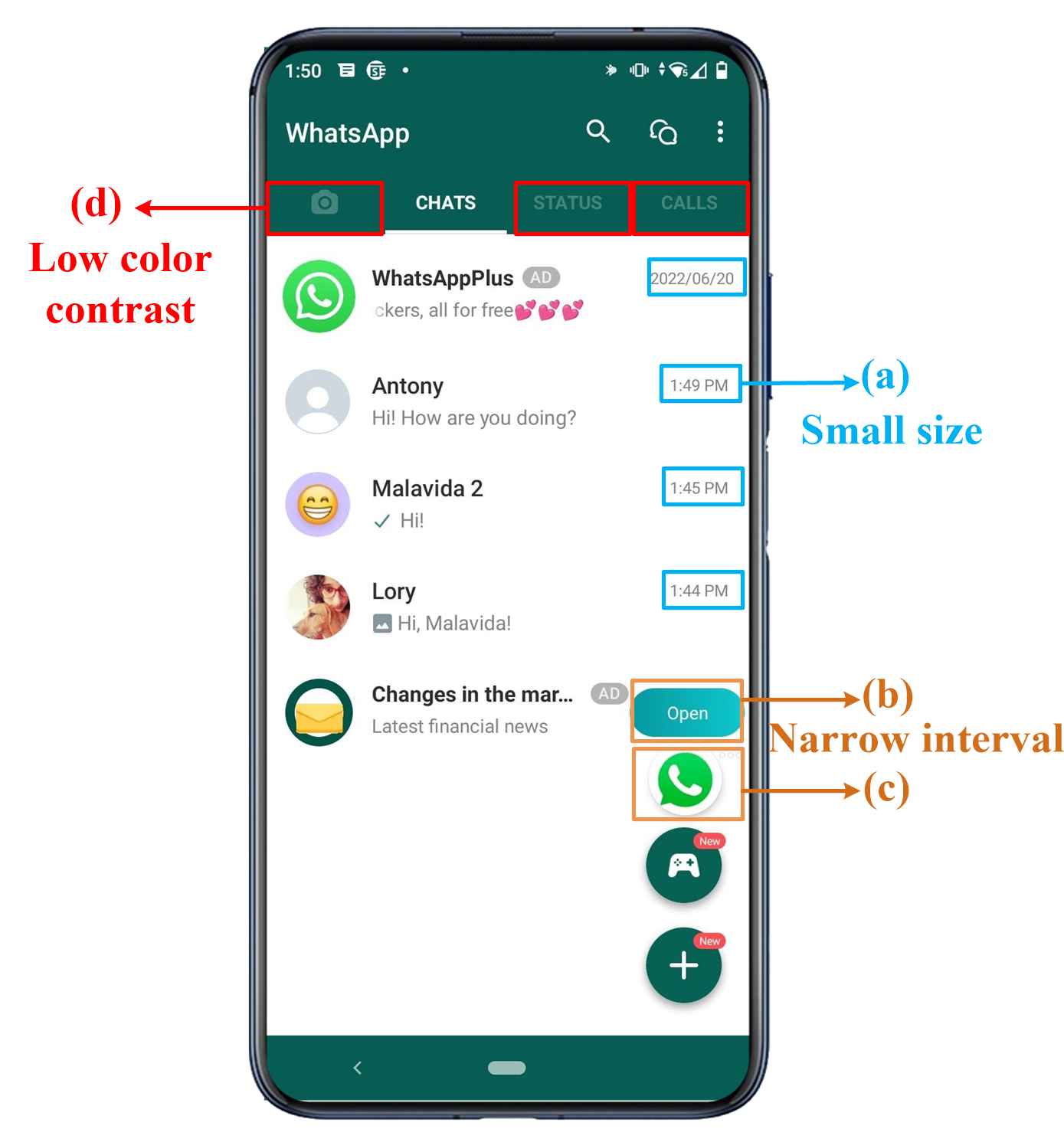}
\caption{The practical example of Whatsapp.}
\label{fig: whatsapp}
\end{figure}

\section{Preliminary research}\label{sec: background}
In this section, we conduct a preliminary study on the accessibility of apps. 
Specifically, we detailedly illustrate the accessibility issues involved in our work using a practical example, as well as a preparatory experiment to investigate the current state of app accessibility.

Figure~\ref{fig: whatsapp} shows a popular communication app, named \emph{Whatsapp}, which has more than 5 billion downloads with a rating of 4.2 in Google Play.
This app provides various functions, such as messaging, calling, exchanging photos, videos, documents, and voice messages.
People without disabilities can use this app smoothly with their clear eyesight, while for low vision users, it is problematic.
For instance, if a low vision user desires to view the chat time, it can be hard for him/her to find it because the size of button (a) is too small.
Furthermore, the interval between buttons (b) and (c) is narrow, which increases the likelihood of low vision users inadvertently touching button (c) while intending to click on button (b).
As for button (d), its color is relatively light compared to the background, which might cause low vision users to overlook this button.
Confronting these practical accessibility issues, low vision users are bound to encounter various difficulties, which may ultimately restrict their ability to fully utilize the functions of apps. 
\begin{table}\footnotesize
\caption{The existing state of affairs on app accessibility.\label{tab: detection}}
\begin{center}
\renewcommand{\arraystretch}{1.5}
\begin{tabular}{cccc}
\hline
\rowcolor{gray!15} \textbf{Tools}  & \textbf{GUIs \textcolor{gray}{(Coms)}} & \textbf{I\_GUI / P1} & \textbf{I\_Component / P2} \cr
\hline
\rowcolor{gray!3} AS & 2,646 \textcolor{gray}{(43,194)} & 2,149 / 81.23\% & 9,245 / 21.41\% \cr
\rowcolor{gray!15} PUMA & 2,646 \textcolor{gray}{(43,194)} & 2,096 / 79.22\%  & 8,386 / 19.41\% \cr
\rowcolor{gray!3} MAC & 2,646 \textcolor{gray}{(43,194)} & 2,119 / 80.11\% & 8,751 / 20.25\% \cr
\rowcolor{gray!15} MATE & 2,646 \textcolor{gray}{(43,194)} & 1,995 / 75.39\% & 8,547 / 19.78\% \cr
\rowcolor{gray!3} Average & 2,646 \textcolor{gray}{(43,194)} & 2,089 / 78.95\% & 8,732 / 20.21\% \cr
\hline
\end{tabular}
\end{center}
\end{table}
\begin{figure*}
\centering
\includegraphics[width=17.5cm]{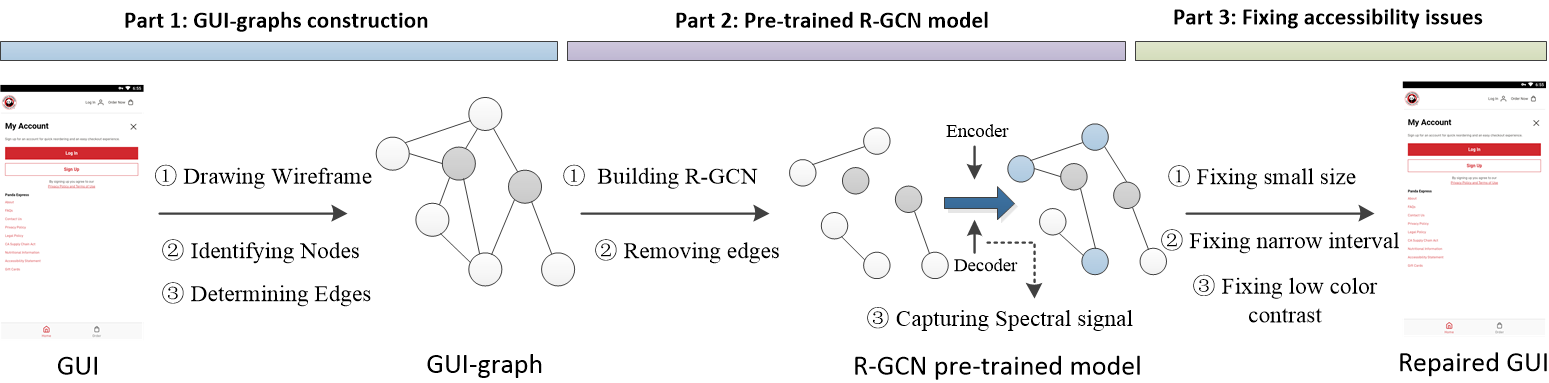}
\caption{The overview of AccessFixer.}
\label{fig: overview}
\end{figure*}

In existing research, there are many tools capable of detecting accessibility issues in GUIs~\cite{AccessibilityScanner}~\cite{Hao2014PUMAPU}~\cite{Eler2018AutomatedAT}. 
Inspired by these efforts, we desire to gain insights into the current state of app accessibility.
As follows, we carry out a preliminary experiment to explore this question.

We start by collecting the Top-500 downloaded apps in Google Play on June 7, 2022.
The reason for selecting such apps is that they have high downloads and are more likely to be commonly used by low vision users.
Also, the design of these apps is relatively optimal, and if these apps suffer from many accessibility issues, then the other apps may also exhibit similar or even worse accessibility results with a high probability.
After that, we adopt an automated testing tool for apps, called App Crawler~\cite{AppCrawler}, whose main function is to simulate users' interactive behavior in apps. 
In our work, with the help of its simulated interaction function, we execute this tool to output the GUI screenshots and layout files (.xml) in apps. 
Then, for tools capable of dynamic analysis (AS and MAC), we perform real-time execution and accessibility detection on these GUIs, while for static tools (PUMA and MATE), the GUI screenshots and files serve as their inputs to detect the accessibility issues.
In total, we obtain 2,646 GUIs covering 43,194 components for this experiment.

After obtaining these GUIs, we use the AS, PUMA, MAC, and MATE to detect their accessibility issues in terms of small size, narrow interval, and low color contrast, respectively.
During this process, some apps may not be successfully identified by the detection tools due to the limitation of the simulator version, so we directly detect them on the physical device to ensure that all GUIs can be detected.

Table~\ref{tab: detection} presents the detection results of existing tools for 2,646 GUIs.
The first column represents the different detection tools, and the second column is the total number of GUIs these tools detect.
The third column is the number of proportion (P1) of GUIs that have accessibility issues (I\_GUI), as well as the number of proportion (P2) of components that have issues (I\_Component) shown in the last column.
The last row shows the average value of each column.
As we can see, an average of 78.95\% of the GUIs contain accessibility issues, as reflected in 20.21\% (8,732 components) of all components are problematic.
This is a disappointing outcome and serves as a clear indication that developers may not have prioritized accessibility as much as they should have done. 
It also potentially indicates a lack of understanding of developers on how to effectively fix the accessibility issues within GUIs.

Focusing on these detection tools, they play a vital role in identifying and locating accessibility issues within the GUIs, but they often fall short of providing solutions to guide developers to fix these issues effectively. 
This limitation might result in a significant gap between issue detection and issue resolution.
It leaves the burden of manually fixing the problems to developers, which may require specialized knowledge and additional time.
AccessFixer, in contrast, aims at overcoming this disparity and surpassing the limitations of existing methods that can only provide issue reports.
In practice, our method can automatically adjust problematic components' attributes to enhance accessibility, thereby providing developers with concrete and practical solutions.
Such problematic components exist in the GUIs when they are associated with accessibility issues of small size, narrow interval, or low color contrast.

\section{AccessFixer: Accessibility Issues Fixing Method}\label{sec: ARBot}

In this section, we introduce our proposed AccessFixer in detail.
This method aims to provide the attributes of GUI components with concrete adjustments, which could be roughly divided into three parts, as seen in Figure~\ref{fig: overview}.
Given a GUI, AccessFixer first draws a wireframe of its components and containers, followed by node identification and edge determination, and this GUI can then be converted into a GUI-graph.
Then, we construct R-GCN pre-trained model, and establish its encoder, decoder, and loss function.
Subsequently, for a GUI to be fixed, we adopt the AS to detect its accessibility issues, and convert it into a GUI-graph.
We further remove the edges connected to the problematic nodes, and then utilize the pre-trained model to predict the connections.
During the prediction process, AccessFixer automatically captures the feature representation of node information within graphs (spectral signal in Figure~\ref{fig: overview}), and then formulates strategies for fixing accessibility issues.
\begin{figure}
\centering
\includegraphics[width=7.5cm]{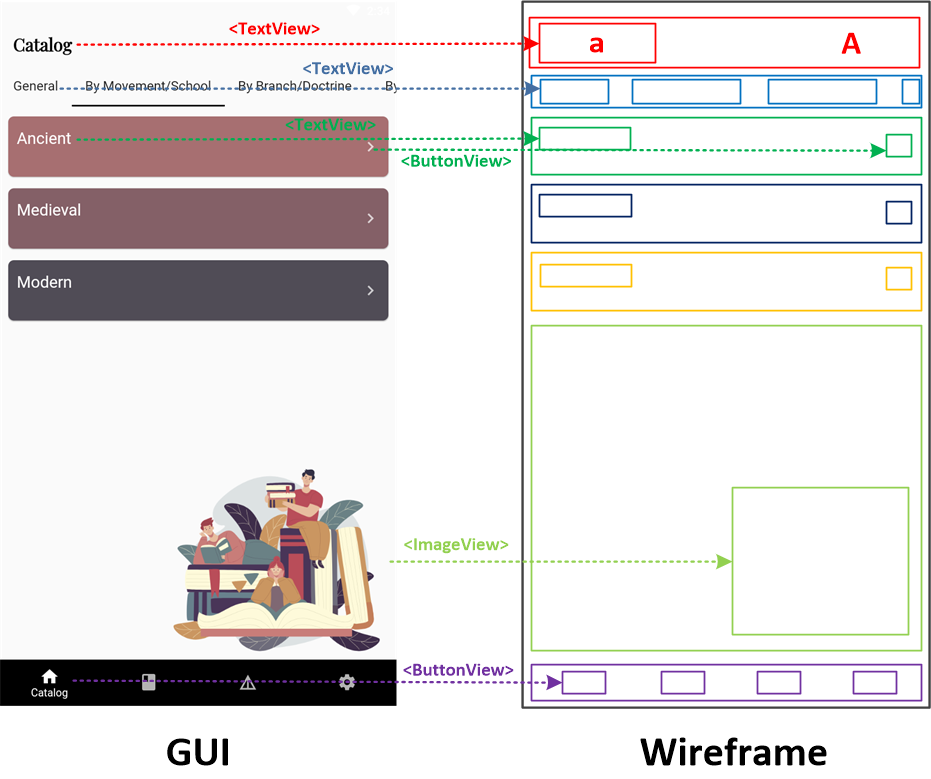}
\caption{An example of drawing wireframe of GUIs.}
\label{fig: wireframe}
\end{figure}
\subsection{GUI-graph construction}\label{sub: formal}

Formalizing GUIs should incorporate both performance and structural features from the components~\cite{Xu2021GUIDERGS}.
Thus, in this part, we represent the GUIs in the form of GUI-graphs according to their performance attributes of positions, sizes, and hierarchy.
In more detail, the aforementioned conversion process can be accomplished through the following three steps.

\subsubsection{Drawing the wireframe structures of GUIs} 

During the process of GUI design, developers commonly use multiple views to standardize and constrain the layout of components, resulting in nested containers and components~\cite{Designprinciples}.
However, for both sighted and low vision users, the information presented in the GUIs is not the internal hierarchy, but the visual information, such as buttons, texts, and images.
Thus, converting such visible information into wireframe structures could effectively present the GUI without considering complex nested layouts.
Moreover, this way is also beneficial for accurately locating visible components and containers in the subsequent construction of GUI-graphs.

To draw the wireframes of GUIs, we adopt the UIAutomator~\cite{UIautomator} to parse GUIs.
The layout files (\emph{.xml}) parsed in this way allow us to focus on two aspects, one is the components within GUIs, and the other is the containers that hold these components.
Regarding the components, they generally exist in three categories of views, one is the \emph{$<$ButtonView$>$}, the next one is the \emph{$<$TextView$>$}, and the last category is the \emph{$<$ImageView$>$}.
In general, we could draw their wireframes directly through the \emph{bounds} in the layout file.
Still, there might be a special case where a component covers or contains another one.
Thus, we further determine the specific positional information among these components when drawing the wireframes.
Notably, components might also be nested by multiple invisible layouts, but we only adopt the innermost structure to present the components since those invisible layouts would increase the complexity of the structure.
The containers normally appear as the \emph{$<$ViewGroup$>$} in the layout files, meanwhile, they might be nested with other containers.
However, these nested containers are not in our consideration because they could not convey any information when drawing wireframes.
Then, we locate the attribute of \emph{bounds} from the layout files, obtain the corresponding left and right vertex coordinates, the length and width, and further draw their wireframes.
The upper left corner of Figure~\ref{fig: wireframe} presents an example of the wireframe we have drawn, like the component $a$ and container $A$.
\begin{figure}
\centering
\includegraphics[width=8.5cm]{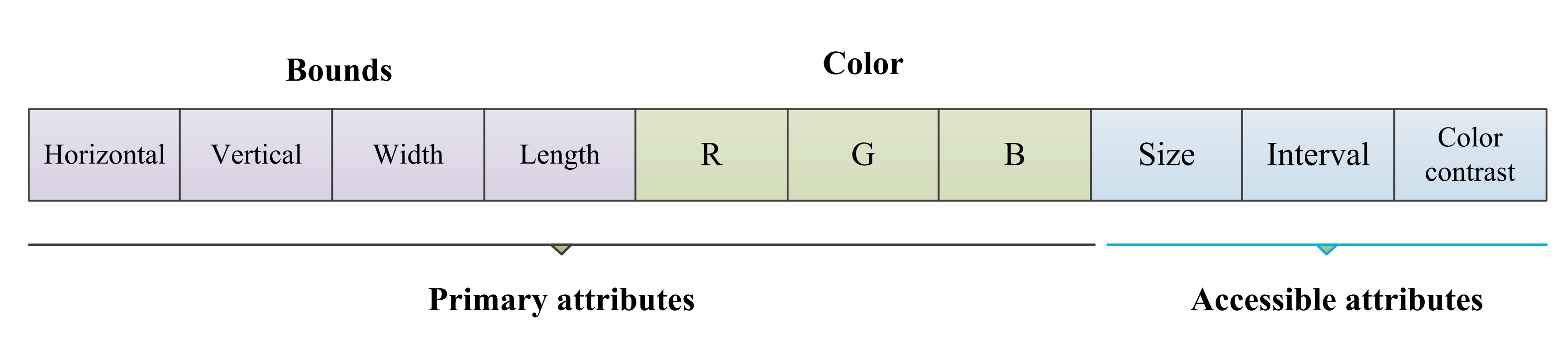}
\caption{The attributes of component nodes.}
\label{fig: vector}
\end{figure}
\begin{figure}
\centering
\includegraphics[width=7.5cm]{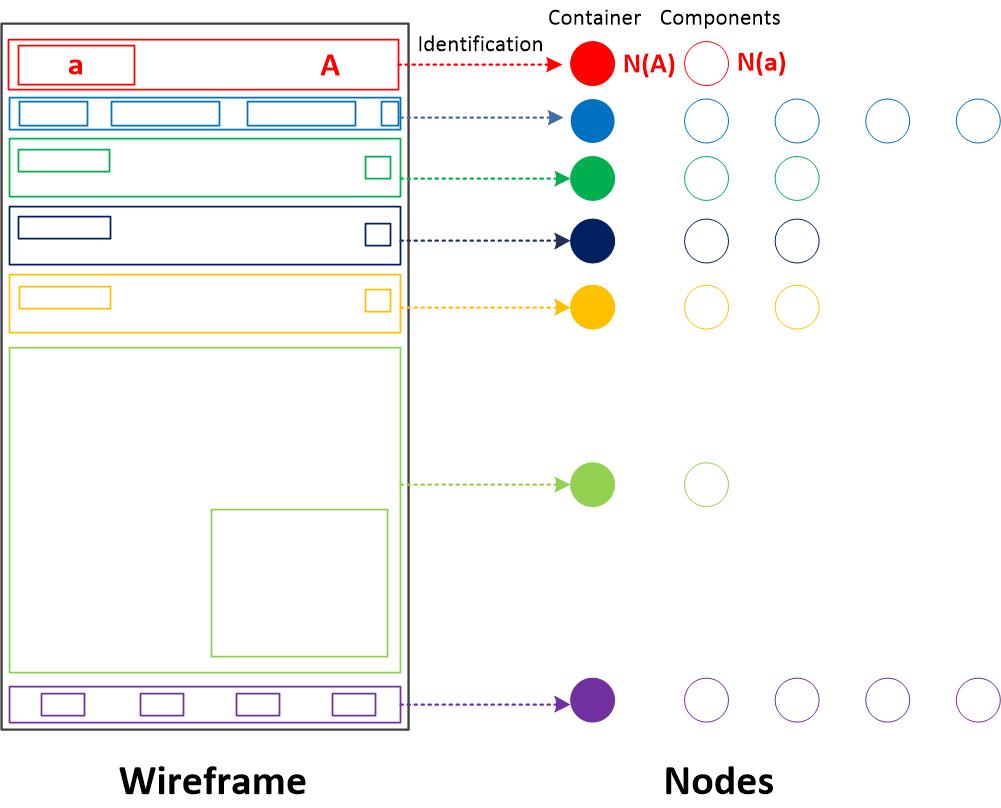}
\caption{An example of identifying the nodes.}
\label{fig: nodes}
\end{figure}

\subsubsection{Identifying the nodes from the wireframe structures}\label{sub: nodes}
In the wireframes, two types of nodes need to be identified, one is the \emph{container-node} and the other is the \emph{component-node}.
For the \emph{container-nodes}, they represent the containers illustrated in the wireframes.
This type of node does not contain any attributes, as it serves solely to divide the GUI into areas and organize other elements.
Regarding the components in the wireframe, each of them is identified as a \emph{component-node}.
Figure~\ref{fig: vector} shows the attributes we add to this type of node, including two kinds of primary attributes and three accessible attributes.
The primary attributes are \emph{``bounds''} and the \emph{``color''} in the \emph{component-node}.
The \emph{``bounds''} occupies 4 bytes that consist of the coordinates of left vertex in the component, and its width and length, which is parsed by UIAutomator~\cite{UIautomator}.
The \emph{``color''} is the value of RGB in components, which contains 3 bytes and is extracted by AirTest IDE~\cite{AirTestIDE}.
The accessible attributes consist of \emph{``size''}, \emph{``color contrast''}, and \emph{``intervals''}.
The \emph{``size''} and \emph{``intervals''} can be determined by computing the length, width, and the difference of vertex coordinates of the \emph{``bounds''}, while \emph{``color contrast''} can be determined by calculating the ratio between the component \emph{``color''} and the background color.
Figure~\ref{fig: nodes} presents an example of our identified nodes from wireframe structures, like node $N(a)$ from component $a$ and node $N(A)$ from container $A$.

\begin{figure}
\centering
\includegraphics[width=7.5cm]{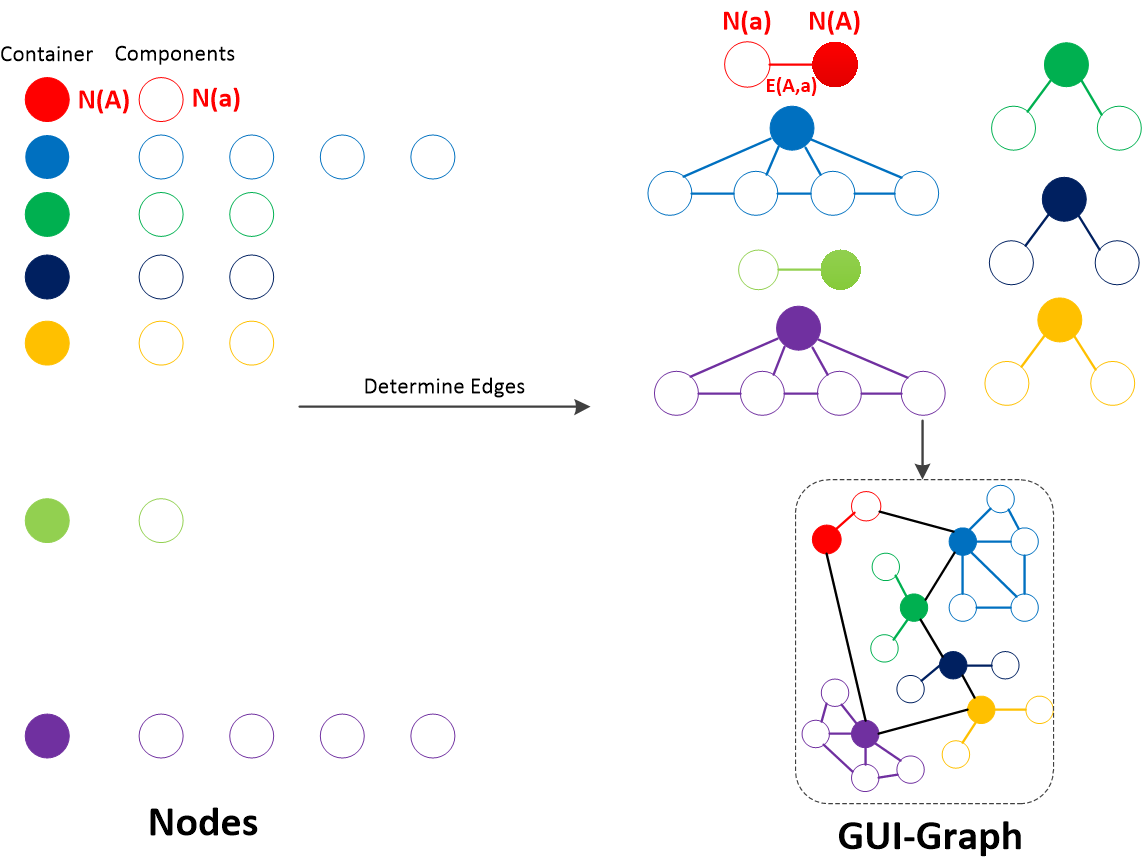}
\caption{An example of determining the edges.}
\label{fig: edges}
\end{figure}
\subsubsection{Determining the edges between nodes}
We now present the three types of edges that may exist between the nodes identified in Section~\ref{sub: nodes}.
The first category is the edge between \emph{component-nodes}.
We draw this category of edge by traversing all \emph{component-nodes} within the same container and connecting adjacent components. 
For instance, container $M$ has four components, namely $C_1$, $C_2$, $C_3$, and $C_4$, which are arranged in a square shape with four vertices. 
In this arrangement, $C_1$ is adjacent to $C_2$, $C_1$ is adjacent to $C_3$, $C_2$ is adjacent to $C_4$, and $C_3$ is adjacent to $C_4$, with each pair of adjacent components connected by an edge.
The second is the edge between the \emph{component-node} and the \emph{container-node}, as reflected in the \emph{container-node} would connect all the \emph{component-nodes} it contains.
The last category is the edge between \emph{container-nodes}.
In this process, all \emph{container-nodes} are connected sequentially based on the positional layout from top to bottom and left to right, and then connect the first \emph{container-node} and the last one.
As shown in Figure~\ref{fig: edges}, it illustrates the results that are ultimately drawn after determining the edges.
An example can be found in the edge $E(A,a)$ that connects the node $N(A)$ and node $N(a)$.

With all nodes identified and edges determined, we construct the GUI-graph of each GUI, supporting further process and the training of R-GCN model in the following section.
In the subsequent training and testing of the model (in Section~\ref{sub: make}), we randomly remove 3 edges of any type in the GUI-graphs.
When using this pre-trained model to fix accessibility issues (in Section~\ref{sub: fixing}), we remove all edges that are connected to the problematic components.
Specifically, the removal of these edges entails changes in the adjacency matrix corresponding to the GUI-graphs.
For instance, to remove an edge between node $i$ and node $j$, it suffices to set the elements located at the $i_{th}$ row and $j_{th}$ column, as well as the $j_{th}$ row and $i_{th}$ column, in the adjacency matrix to 0.

\begin{figure*}
\centering
\includegraphics[width=17.5cm]{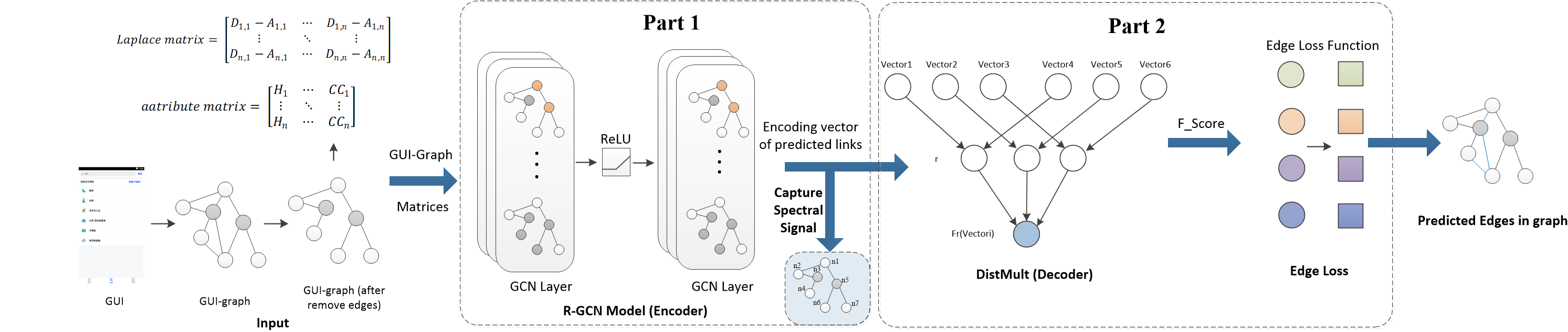}
\caption{The architecture of our pre-trained model based on R-GCN.}
\label{fig: R-GCN}
\end{figure*}
\subsection{Pre-trained model based on R-GCN}\label{sub: RGCN}

In this part, we first show the rationale behind adopting the R-GCN model for rectifying accessibility issues, together with introducing and building the architecture of our R-GCN pre-trained model.
In our model, we explain a method to extract the feature representation of node information within graph structures, to provide technical support for fixing accessibility issues.
Finally, we make the dataset to train this model, and obtain a trained model that has learned the features of the well-accessible GUIs. 

\subsubsection{Explanation of the pre-trained model}
As a branch of the Graph Convolutional Network (GCN), R-GCN could combine relational data modeling with the GCN framework.
Utilizing this model to fix accessibility issues in GUIs primarily stems from the complex relationships and interactions among the GUI components.
Specifically, the position and layout of one component may affect the arrangement of other components, while alterations in attributes (i.e., size, interval, and color) necessitate related adjustments in other components to ensure a consistent color palette, uniform intervals, and adequate sizes.
The robust relational processing capability of the R-GCN model enables it to effectively capture these inter-component relationships, facilitating accurate processing. 
Also, this model possesses a mechanism for adaptively adjusting the weight parameters and sharing them when predicting links~\cite{Schlichtkrull2017ModelingRD}.
Such a mechanism allows nodes to leverage shared weights for relational modeling, which breaks away from using the fixed weights in traditional graph neural networks.
It also enables the model to adjust the contribution of neighboring node features based on each node's characteristics.
In parallel, this mechanism is concomitant with the ongoing evolution of the spectral signals of each node within the graph, persisting until the model anticipates all connections and attains stability~\cite{Schlichtkrull2017ModelingRD}.

Figure~\ref{fig: R-GCN} shows the basic architecture of our constructed pre-trained model, consisting of the contents within two parts (Part 1 and Part 2).
Part 1 represents the encoder of the model, while Part 2 comprises the model's decoder and loss function.
When training and testing the model, we input GUI-graphs where some edges have been randomly removed, along with their corresponding adjacency matrix and attribute matrix. 
The output is the model's prediction of the missing edges after the completion of Part 2.
Meanwhile, during model testing, we also capture and record the spectral signal values for each node, to establish mapping sets between the attributes of each component and these values, as described in Section ~\ref{sub: make}.
In the process of using this model, we provide it with a GUI-graph in which edges connected to problematic components have been removed. 
Our focus then shifts to Part 1, from which we extract and output the spectral signal values generated by the model for each node once stability is reached.
Regarding this signal, we provide detailed explanations in the following content.

Subsequently, we explain the details of each content in the architecture built in Figure~\ref{fig: R-GCN}.
The input we constructed is to initialize and process the GUI-graphs, ensuring that the model has sufficient learning capacity during the training process~\cite{He2015DelvingDI}.
This step involves computing and normalizing the Laplacian matrix~\footnote{The Laplacian matrix is defined as $D-A$, where $D$ is a diagonal matrix with the degree of each node, and $A$ is the adjacency matrix.} of GUI-graphs, together with combining the attributes of \emph{component-nodes} in the GUI-graphs into the attribute matrix.
Our constructed attribute matrix comprises the attributes of all component nodes within the GUI-graphs, encompassing both the ``primary attributes'' and ``accessible attributes'' as depicted in Figure~\ref{fig: vector}. 
In this matrix, each row corresponds to a node, while each column corresponds to a specific attribute.
In terms of the R-GCN model (Encoder in Part 1), it consists of the graph convolutional layer, the activation function, and the pooling layer.
Among them, we establish two convolutional layers, ReLU activation operation and max-pooling layers in the R-GCN model.
We choose to use two convolutional layers according to our test on the performance of the model using one to ten layers independently.
We find that the model achieved the optimal performance with two convolutional layers.
If we add more layers, it would cause gradient descent and result in a decrease in the accuracy of the model~\cite{He2015DeepRL}.
Besides, after comparing the performance of the model under mean-pooling and max-pooling, we select the max-pooling principle with higher predictive ability.

Following this, the model generates the encoding vectors for predicted links and feeds them into the decoder to produce embedded representations for each node.
The embedded representations of the node features are viewed as spectral signals, which lie in a transformed space that captures the contextual relationships between nodes, and is defined as the feature projection of nodes or edges into a low-dimensional space~\cite{Schlichtkrull2017ModelingRD}.
These signals encapsulate crucial information about the interactions and dependencies among nodes in the graphs, enabling the model to comprehend the patterns of neighboring nodes and their connectivity. 
In our work, when utilizing the pre-trained model for fixing accessibility issues in Section~\ref{sub: fixing}, we solely focus on the variation of spectral signal and do not engage in link prediction.
This is because the signal value can be correlated with the attributes of nodes, and the signal of each node will undergo variation as the number of iterations increases until the model reaches its optimal state.
In this manner, our approach leverages signal values associated with nodes representing GUI components to fix problematic components.
Also, the signals of other nodes would co-evolve to maintain a consistent color palette, uniform intervals, and adequate sizes in GUIs.

We now describe the detailed process of capturing the spectral signal, which typically involves two stages of signal processing and spectral analysis~\cite{Kipf2017SemiSupervisedCW}.
In the signal processing stage, as the time for each node to reach a stable state is different, we need to continuously monitor the signal changes of all nodes. 
Therefore, we do not conduct any pre-processing such as filtering and gaining control of the original signal in this stage.
During the stage of spectral analysis, we use the Fourier transform~\cite{bracewell1986fourier} (formula~\ref{for: fouriert}) to convert the signal of the time domain into the frequency domain.
We then extract the transformed Fourier coefficients (formula~\ref{for: forierc}) as the spectral signal values of specific nodes.
The reason for doing this is that the Fourier coefficients provide a quantitative description of signal fluctuations in the frequency spectrum~\cite{Vaibhav2018NonlinearFT}.
\begin{equation}
\label{for: fouriert}
\hat{f}(\xi)=\int_{-\infty}^{\infty}f(x)e^{-2\pi ix\xi}dx
\end{equation} 
where $\hat{f}(\xi)$ refers to the representation of function $f(x)$ in the frequency domain, and $\xi$ is the frequency.
\begin{equation}
\label{for: forierc}
c_k=\frac{1}{N}\int_{n_0}^{n_0+N}f(n)e^{-i2\pi n\frac{n-n_0}{N}}dn
\end{equation}
where $c_k$ represents the complex Fourier coefficient of the $k$-th frequency component, $f(n)$ represents the sampling value of the $n$-th sampling point, and $N$ represents the number of sampling points.
With this approach, we obtain the spectral signal of nodes, and accordingly, construct the mapping set between these signal values and the attributes of components in Section~\ref{sub: make}.

As for the Part 2 in our model, it includes the DistMult~\cite{Yang2015EmbeddingEA} as the decoder~\cite{Li2020AutoCO}, and the edge loss function.
Regarding the DistMult~\cite{Yang2015EmbeddingEA} (Decoder), it aims to evaluate the accuracy of link prediction via a scoring function.
The scoring function is defined by referring to the work~\cite{Wang2017KnowledgeGE}, and it is set as the inner product of two node vectors.
We regard the highly integrated DistMult~\cite{Guo2017KnowledgeGE} as the decoder, since it can effectively handle large-scale graph structures, and can reduce the overfitting problem~\cite{Vashishth2019CompositionbasedMG}.
The last one is the edge loss function to determine when the training model should stop iterating, and identify the target connections.
For this, we refer to the union loss function~\cite{Schlichtkrull2017ModelingRD} of predicting edges shown in the following formula~\ref{for: loss}, while limiting the model to a stable condition and ensuring that the results are consistent with the training samples at least.
\begin{equation}
\begin{aligned}
\label{for: loss}
L = -\frac{1}{(1+\omega)|\hat{E}|} \sum_{(s,r,o,y) \in T} y \cdot log(l(f(s,r,o))) \\+ (1-y)\cdot log(1-l(f(s,r,o)))
\end{aligned}
\end{equation}
where $(s,r,o)$ is a triple of graph feature information, and among them, $s$ stands for the subject and starting point of the relationship triple, which is typically an entity or a node, $r$ represents the relationship connecting the subject $s$ and object $o$, and $o$ denotes the object and endpoint that is also an entity or a node.
$f()$ refers to the scoring function for the triple samples.
$\omega$ and $|\hat{E}|$ separately indicate the weight and number of edges.
By following this approach, we can capture the spectral signal values for each node in the graphs.

Overall, compared to the original R-GCN model, there are differences in the dimensions of the data our model processes during the encoding phase.
Therefore, we adjust the parameter of the dimension of the input features to enable the model to handle our data. 
Additionally, in the decoding process, both our model and the original model predict missing edges in the graph. 
However, our method's ultimate focus is not on these edges, but rather on the variations in spectral signal values of individual nodes during the prediction process. 
Overall, our model shares similarities with the original R-GCN model, except for the data contents and the specific outputs.

\subsubsection{Making the training sets and test set}\label{sub: make}
Due to the fact that we aim to use the constructed model for fixing accessibility issues in GUIs, the dataset utilized for training the model comprises GUIs without any accessibility issues.
To this end, we first collect the apps with top-300 downloads from each of the six domains (\emph{Communication}, \emph{Entertainment}, \emph{Music}, \emph{Book}, \emph{Sociality}, and \emph{Shopping}) in Google Play, and obtain 8,554 GUIs using App Crawler~\cite{AppCrawler}. 
Then, we evaluate these GUIs through AS, PUMA, MATE, and Xbot, aiming to detect accessibility issues within them and select those without issues detected by any of the tools. 
In total, 2,050 GUIs conform to this standard. 
Further, among these GUIs, we make a second evaluation with real low vision users. 
Therefore, we recruit 20 users (10 males and 10 females) from the Special Education College to observe these GUIs, and ask them to tell us which GUIs can be seen clearly and which are not. 
In this process, each GUI is evaluated by 2 users, and these users will not participate in any subsequent experiments after completing the collection of the dataset to avoid subjective bias. 
To accommodate this, we provide these users with two weeks to complete the task without disrupting their daily routines.
Each user is assigned an average of 205 GUIs, and they can do this task during their idle time. 
After two weeks, all users complete the task, and we further conduct two rounds of manual validation of their provided results to ensure the selected GUIs without any uncertainties. 
In total, we collect 1,925 available GUIs, with 1,540 GUIs (80\%) for training and 385 GUIs (20\%) for testing.
Notably, in the test sets, we randomly remove 3 edges in each case with a total of 1,155 removed edges, allowing the model could re-predict the links and conduct the test.

After testing 385 GUIs without any accessibility issues in our test set and recording their signal values, we create mapping sets between the attributes of each component and its stable spectral signal.
We construct this mapping because it allows us to determine the value of the attribute to be adjusted for fixing accessibility issues based on the signal values. 
For instance, regarding a \emph{component-node} $I$, its signal value when reaching stability is $D_1$.
Regarding the size of component $I$, it can be expressed as $S$, then there is a mapping $(S,D_1)$ between the size and stable signal value.
There is another \emph{component-node} $J$ whose interval from $I$ is denoted as $V$.
The stable signal value of component $J$ is $D_2$, indicating there is a mapping between the signal values and the interval of the two components, which is represented as $(V,|D_1-D_2|)$.
As for the color contrast of $I$ and the background called $CR$, we recall that it is the ratio calculated by the RGB color of components and the background, and there is a mapping of $(CR,D_1)$.

\subsection{Fixing the accessibility issues in GUIs}\label{sub: fixing}
With the well-constructed mapping sets, the stable signal value of each component is related to the attributes of size, interval, and color contrast.
Meanwhile, different components exhibit distinct mapping relationships, and the data distribution appears non-linear. 
To unify these relationships, we adopt a polynomial regression model~\cite{Polynomialregression} to fit each type of mapping.
This fitting can be applied to any complex non-linear function, enabling us to precisely identify the attribute values corresponding to the stable signal values of the components when fixing accessibility issues.
For each type of accessibility issue, we separately discuss the calculation method and possible scenarios involved.
Also, once there are multiple accessibility issues in a component, each issue needs to be resolved, respectively.

\subsubsection{Locating the problematic components} 

To figure out where the problematic components are in the GUIs, we adopt the Google Accessibility Scanner~\cite{AccessibilityScanner} to check the GUIs.
This tool can annotate accessibility issues in GUIs (e.g., the lack of content-description and small target sizes), and provide text alerts for users~\cite{AccessibilityScannerAndroid}.
We then select the components signed with small sizes, narrow intervals, and low color contrast from the results.
In this process, the detection results of this scanner are presented on the GUI screen, which means its contents cannot be directly used.
Thus, we adopt Appium~\footnote{An open-source test automation framework for mobile apps. It allows testers to write automated tests for native and hybrid.} to save and record the results on the GUI screens into text.

\subsubsection{Fixing the accessibility issues} 
For a GUI to be fixed, our approach first locates its problematic components using the method described above. 
Then, all edges connected to these nodes are removed from the corresponding GUI-graphs, and the resulting graph is inputted into our pre-trained model. 
AccessFixer captures the spectral values of all nodes as the model predicts the connections of the inaccessible nodes to other nodes.
The signal value of each node will converge at different iteration rounds until all connections are successfully predicted by the model, and the iteration stops. 
Referring to the stable signal values of these nodes and the polynomial fitting curve we obtained by combining the mapping relationship (following formulas 4,5, and 7), our method can adjust the attributes of all components while simultaneously fixing the three types of accessibility issues.
Such fitting functions rely on the mapping set constructed in Section~\ref{sub: make}, where the relationships between each type of issue and its corresponding signal are individually fitted into quadratic functions using polynomial regression.
Notably, the coefficients correspond to the various feature terms in the fitted curve, which control the shape and trend of the curve.

\begin{figure}
\centering
\includegraphics[width=8cm]{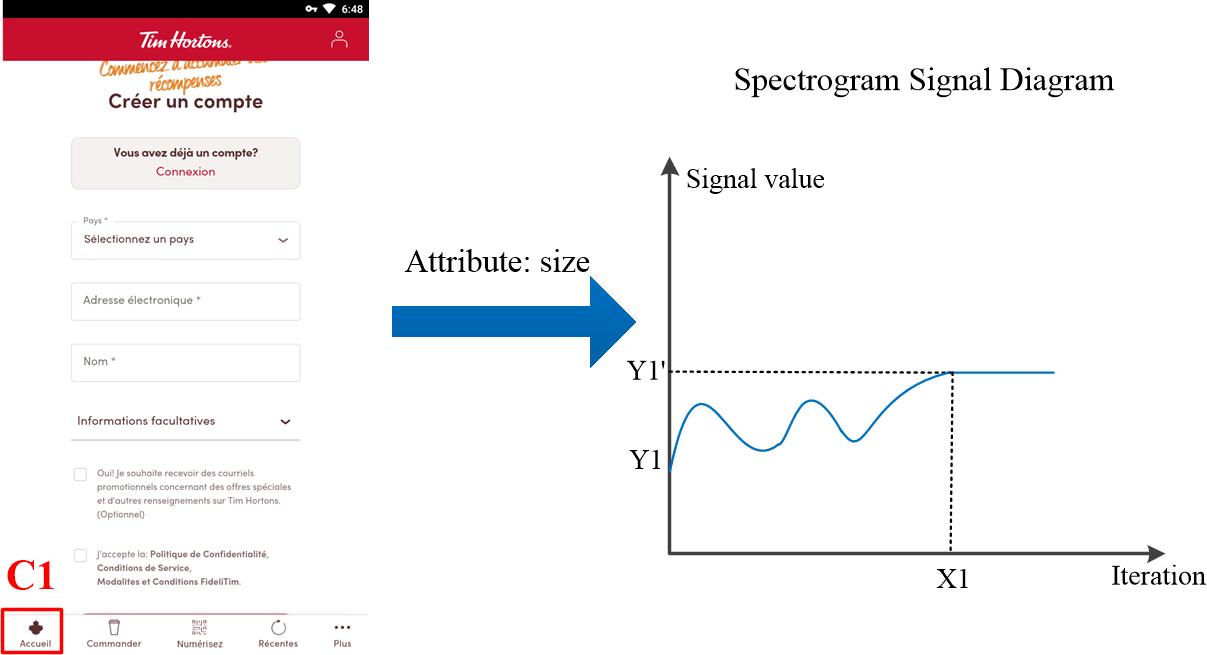}
\caption{An example of fixing small size issue.}
\label{fig: size}
\end{figure}

\emph{1) Fixing accessibility issue of small size:} Given a GUI component that has such issues, the fixed principle follows the mapping set we constructed, and its polynomial fitting result is shown in formula~\ref{for: for1}.
\begin{equation}
\small
\label{for: for1}
f_{size}(\alpha) = 0.045\alpha^2+1.742\alpha-0.0256
\end{equation}
where $\alpha$ indicates the signal value, and $f_{size}(\alpha)$ refers to the value of size corresponding to this signal value.
An example is shown in Figure~\ref{fig: size}, where component $C1$ is detected to have the issue of small size.
After this GUI is fed into our pre-trained model, the signal value change of component $C1$ is shown on the right of this figure.
Thereinto, $Y1$ represents the signal value for the original size of $C1$, $Y1'$ refers to the signal value when the model reaches a stable state, and $X1$ represents the number of iterations required for this component to stabilization.
This way, the value of $f_{size}(Y1')$ is the component $C1$ that needs to be resized.
\begin{figure}
\centering
\includegraphics[width=9cm]{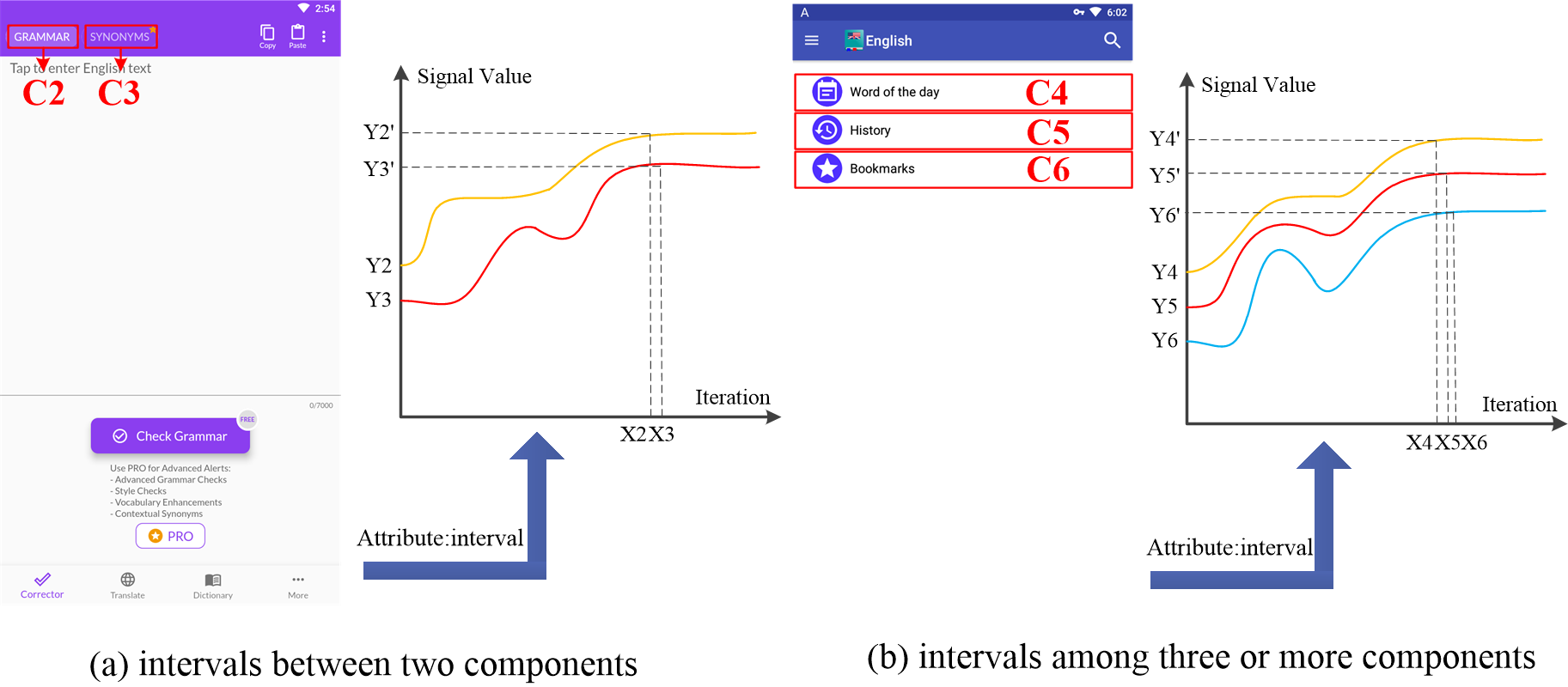}
\caption{Examples of fixing narrow interval issue.}
\label{fig: interval}
\end{figure}

\emph{2) Fixing accessibility issue of narrow interval:} If there are narrowly-spaced components in the input GUI, referencing the fitting curve shown in formula~\ref{for: for2} is necessary to fix them.
\begin{equation}
\small
\label{for: for2}
f_{interval}(\alpha_1, \alpha_2) = 0.042(|\alpha_1-\alpha_2|)^2+1.634(|\alpha_1-\alpha_2|)-0.0378
\end{equation}
where $f_{interval}(\alpha_1, \alpha_2)$ indicates the interval size corresponding to the signal values of $\alpha_1$ and $\alpha_2$ in two components.

Meanwhile, it may involve the joint adjustment of multiple components, and two different scenarios require consideration.
One is that there are only two components that need to be adjusted, as seen in the components $C2$ and $C3$ in Figure~\ref{fig: interval} (a).
Feeding this GUI into our pre-trained model, the signal values of two components change iteratively until they reach stability at $X2$ and $X3$ iterations, with their stable signal values being $Y2'$ and $Y3'$, respectively.
Thus, the interval of components $C2$ and $C3$ should be adjusted to $f_{interval}(Y3',Y2')$.
The other scenario involves adjusting the interval between three or more components in the same container.
In this case, components within the same container should have an equal interval, but in our results, there may be a slight deviation in the interval that should be adjusted between components, so we adopt the following formula~\ref{for: for3} to calculate their mean value as the interval.
\begin{equation}
\small
\label{for: for3}
Mean\_I = \frac{1}{k-1}\sum_{m=2}^k f_{interval}(\alpha_m,\alpha_{m-1})
\end{equation}
where $Mean\_I$ is the interval between components, $k$ is the number of components involved, and $m$ is the sequence number in which the components are ordered from left to right or top to bottom in position.
$\alpha_m$ is the signal value of the $m$-th component, and $f_{interval}(\alpha_m, \alpha_{m-1})$ represents the value of interval when the signal values are $\alpha_m$ and $\alpha_{m-1}$ of components.
Taking components $C4$, $C5$, and $C6$ in Figure~\ref{fig: interval} (b) as an example, when our model is stabilized, the signal values for these three components are $Y4'$, $Y5'$, and $Y6'$, respectively.
As such, we can obtain that the interval between two adjacent components of these three components is $[f_{interval}(Y6',Y5')+f_{interval}(Y5',Y4')]/2$.

\begin{figure}
\centering
\includegraphics[width=8cm]{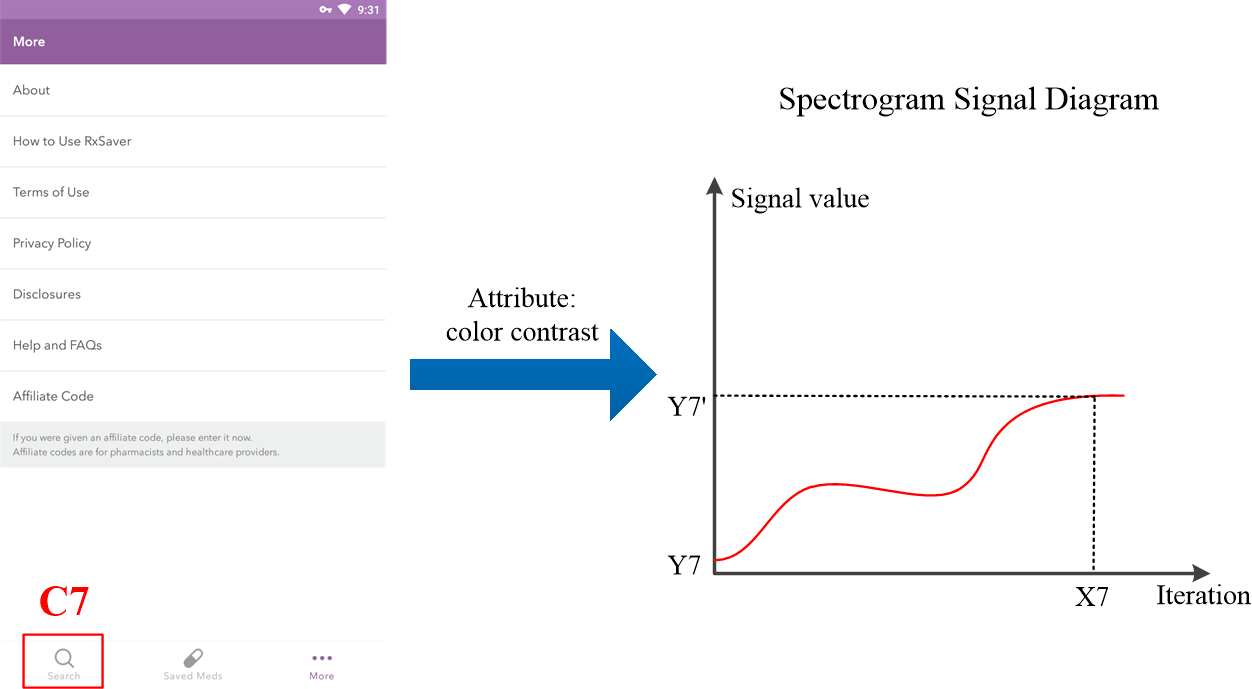}
\caption{An example of fixing low color contrast.}
\label{fig: color}
\end{figure}
\emph{3) Fixing the accessibility issue of low color contrast:} Two schemes might be available for fixing this kind of issue, one is to adjust the color of components, and the other is to adjust the color of the background.
However, after manually adjusting the color contrast of a large number of GUIs, we find that adjusting the background color might lead to a significant alteration in the overall effect of the GUI.
Therefore, our scheme focuses only on adjusting the color of components, while leaving the background color unchanged.

Given a component with low color contrast, it first needs to get the color contrast that should be adjusted with the background according to formula~\ref{for: for4}.
\begin{equation}
\small
\label{for: for4}
f_{color}(\alpha) = 1.328\alpha^2-0.7723\alpha-1.8954
\end{equation}
where $f_{color}(\alpha)$ refers to the value of color contrast at the signal value of $\alpha$. 
Afterward, following the formula~\ref{for: for5}, we calculate the RGB value of this component that should adjust based on the obtained color contrast and the background color~\footnote{\url{https://www.w3.org/TR/WCAG20/\#contrast-ratiodef}}.
\begin{equation}
\small
\label{for: for5}
(R_{cr}, G_{cr}, B_{cr}) = (\frac{L_{bg}+0.05}{cr \cdot 0.2126}, \frac{L_{bg}+0.05}{cr \cdot 0.7125}, \frac{L_{bg}+0.05}{cr \cdot 0.0722})
\end{equation}
where $R_{cr}$, $G_{cr}$, and $B_{cr}$ refer to the color of components, $L_{bg}$ is the brightness of the background, which can be calculated from the RGB value of the background color, and $cr$ is the specific value of $f_{color}(\alpha)$.
Figure~\ref{fig: color} shows an example where component $C7$ has this kind of issue, and its variation in signal value is depicted on the right of this figure.
When the number of iterations reaches $X7$, the model stabilizes, and the signal value of this component becomes $Y7'$.
Then, we can determine that the color contrast of this component and the background color is $f_{color}(Y7')$, and further, the color of this component should be adjusted to $(R_{f_{color}(Y7')}, G_{f_{color}(Y7')}, B_{f_{color}(Y7')})$.

In addition to the problematic components that require adjustment as mentioned above, the signal values of other components in the GUIs also undergo slight changes during the iteration process. 
We also make slight adjustments to their size, interval, and color contrast to ensure the overall visual consistency of the GUIs.

\subsection{Implementation and application}
We implement AccessFixer by packaging the GUI-graphs construction module and the fixing module based on the pre-trained model as an executable file, to support providing adjustments for GUI component attributes.
Our code and datasets have been published and uploaded to \emph{Zenodo}~\footnote{\url{https://doi.org/10.5281/zenodo.7861881}}.

AccessFixer could be executed with the following steps.
Start by providing a GUI to be fixed and its corresponding layout file.
Then, AccessFixer calls the interface of Google Accessibility Scanner~\cite{AccessibilityScanner} to detect the accessibility issues in this GUI, marks the problematic nodes in the constructed GUI-graph, and removes all edges connected to these nodes (in $\mathtt{Graph\_Preparation()}$).
Afterward, this GUI will be fed into our pre-trained model to predict the connections of problematic nodes.
In this process, the spectral signal of nodes is obtained by $\mathtt{Capture\_Signal()}$ in real-time, as well as the signals and attributes are mapped in $\mathtt{Attribute\_Map()}$ to generate the conceret adjustments for each GUI component.
At this point, our method provides specific values that should be adjusted for the attributes within GUI components.
Developers can then utilize such values to update the code, thereby fixing accessibility issues in GUIs.

\section{Evaluation of AccessFixer}\label{sec: evaluation}

In this section, we evaluate our proposed AccessFixer by answering the following two research questions.
\\
\textbf{RQ1 (effectiveness):} Can AccessFixer effectively fix problematic components?
\\
\textbf{RQ2 (usefulness):} How useful is AccessFixer for developers and low vision users?
\\
RQ1 is responsible for elaborating on the effectiveness of AccessFixer, including the quantitative analysis and qualitative discussion of AccessFixer and other baseline tools.
RQ2 discusses whether the fixed GUIs are useful for developers by submitting PRs, and for low vision users through a user study.

\subsection{RQ1: Evaluation of the effectiveness}\label{sub: effectiveness}

To answer this research question, we carry out the following two experiments.
One is to adopt four existing tools to check the accessibility issues of GUIs before and after fixing, and then calculate the percentage of change.
The other experiment focuses on comparing AccessFixer with the baseline tool proposed by Aalotaib et al~\cite{Alotaibi2021AutomatedRO}.

\subsubsection{Data Collection}\label{sub: data}
Evaluating all apps (more than 2.6 million~\cite{GooglePlay}) hosted on Google Play by our approach is a challenging task for us.
Following the existing research~\cite{Bajammal2021SemanticWA}~\cite{Xie2022PsychologicallyinspiredUI}, we attempt to use the subsets of apps to discuss the effectiveness of our method.
Therefore, on June 7, 2022, we randomly collect the apps from the recommendation listed on Google Play, covering 30 apps in six domains (\emph{Communication}, \emph{Entertainment}, \emph{Music}, \emph{Book}, \emph{Sociality}, and \emph{Shopping}).
These 30 apps could be divided into three groups according to their downloads and ratings.
The first group consists of 10 apps, all of which have more than 1 million downloads and their ratings are higher than 4 in Google Play.
The second group also has 10 apps with more than 1 million downloads but all are rated below 4.
While for the third group, it includes the remaining 10 apps with less than 1 million downloads and a rating of below 4.
This grouping allows us to better discuss the effectiveness of AccessFixer for apps with different popularity.
We then automatically intercept GUIs from these 30 apps using App Crawler~\cite{AppCrawler}.
This requires us to install all apps into the Android emulator (Android 11.0 with 16 GB RAM and 2.4GHz Intel i7 9600) and login to these apps in advance, so that App Crawler can run successfully and intercept GUIs.
In total, we collect 284 GUIs from these 30 apps.

Afterward, we construct the ground truth for the experiments.
We first use the AS~\cite{AccessibilityScanner}, PUMA~\cite{Hao2014PUMAPU}, MATE~\cite{Eler2018AutomatedAT}, and Xbot~\cite{Chen2021AccessibleON} to individually detect issues within our dataset of 284 GUIs.
Then, from the detection results, we select components with the issues of small size, narrow interval, and low color contrast, as the ground truth.
We further discuss whether AccessFixer could fix these problematic components in our ground truth effectively.
Besides, as the detection rules established by different tools vary, the number of accessibility issues they can identify also differs.

\subsubsection{The effectiveness of AccessFixer}\label{sub: effectiveness1}

In this experiment, we first use AccessFixer to fix the problematic components in our ground truth.
However, since the apps are closed-source, we cannot fix the accessibility issues by tweaking their source codes directly.
Therefore, we employ the method proposed by Moran et al.~\cite{Moran2018MachineLP} to convert the GUIs of these apps into executable codes, and then fix the problematic components in such codes.
This kind of re-generated codes could well restore the prototyping of GUIs, so it is enough to support our experiments.
In addition, we use Androguard~\cite{Androguard} to decompile these apps and manually inspect the implementation of each view to ensure the generated contents are consistent with the apps.
Then, we render and package the fixed GUIs into installable \emph{.apk} files.

Subsequently, we install these \emph{.apk} files into the Android emulator, and re-check them using the AS, PUMA. MATE, and Xbot, respectively. 
We record the number of problematic components in the fixed GUIs, and compare them with the results before fixing.
\begin{table}[t]
\renewcommand{\arraystretch}{1.4}
\tabcolsep=0.07cm
\caption{Effectiveness of AccessFixer on 30 real-world apps.} \label{tab: effectiveness}
\begin{center}
\begin{tabular}{l|cccc|c}
\hline
\textbf{apps} \textcolor{gray}{(GUIs)} & \textbf{AS} & \textbf{PUMA} & \textbf{MATE} & \textbf{Xbot} & \textbf{Ratio}\cr 
\hline
\cellcolor{gray!8}1. Whatsapp Messenger \textcolor{gray}{(11)} & \cellcolor{gray!8}15/4  & \cellcolor{gray!8}10/2 & \cellcolor{gray!8}8/3 & \cellcolor{gray!8}10/3 & \cellcolor{gray!8}0.72 \cr
\cellcolor{gray!3}2. Facebook \textcolor{gray}{(12)} & \cellcolor{gray!3}11/3 & \cellcolor{gray!3}11/2 & \cellcolor{gray!3}9/2 & \cellcolor{gray!3}12/1 & \cellcolor{gray!3}0.81 \cr
\cellcolor{gray!8}3. Messenger Lite \textcolor{gray}{(8)}  & \cellcolor{gray!8}12/4 & \cellcolor{gray!8}9/3& \cellcolor{gray!8}7/2 & \cellcolor{gray!8}11/3 & \cellcolor{gray!8}0.69 \cr
\cellcolor{gray!3}4. Android Auto \textcolor{gray}{(7)} & \cellcolor{gray!3}9/1 & \cellcolor{gray!3}10/2 & \cellcolor{gray!3}9/2 & \cellcolor{gray!3}10/1 & \cellcolor{gray!3}0.85 \cr
\cellcolor{gray!8}5. TrackSolid \textcolor{gray}{(13)} & \cellcolor{gray!8}17/5 & \cellcolor{gray!8}12/3 & \cellcolor{gray!8}11/4 & \cellcolor{gray!8}15/4 & \cellcolor{gray!8}0.73 \cr
\cellcolor{gray!3}6. Telegram \textcolor{gray}{(11)} & \cellcolor{gray!3}9/2 & \cellcolor{gray!3}9/2 & \cellcolor{gray!3}8/1 & \cellcolor{gray!3}9/1 & \cellcolor{gray!3}0.85 \cr
\cellcolor{gray!8}7. File Miner \textcolor{gray}{(9)}& \cellcolor{gray!8}12/2 & \cellcolor{gray!8}11/2 & \cellcolor{gray!8}10/2 & \cellcolor{gray!8}12/2 & \cellcolor{gray!8}0.82 \cr
\cellcolor{gray!3}8. Keepclean \textcolor{gray}{(8)} & \cellcolor{gray!3}19/4 & \cellcolor{gray!3}15/2 & \cellcolor{gray!3}15/2 & \cellcolor{gray!3}18/4 & \cellcolor{gray!3}0.82 \cr
\cellcolor{gray!8}9. ClassDojo \textcolor{gray}{(11)} & \cellcolor{gray!8}11/3 & \cellcolor{gray!8}7/1 & \cellcolor{gray!8}6/1 & \cellcolor{gray!8}13/3 & \cellcolor{gray!8}0.78 \cr
\cellcolor{gray!3}10. Twitter \textcolor{gray}{(14)} & \cellcolor{gray!3}10/2 & \cellcolor{gray!3}8/2 & \cellcolor{gray!3}8/2 & \cellcolor{gray!3}11/3 & \cellcolor{gray!3}0.71 \cr
\cellcolor{blue!8}11. Shell Asia \textcolor{gray}{(9)} & \cellcolor{blue!8}12/2 & \cellcolor{blue!8}10/2 & \cellcolor{blue!8}9/2 & \cellcolor{blue!8}12/2 & \cellcolor{blue!8}0.81 \cr
\cellcolor{blue!3}12. Shopee \textcolor{gray}{(11)} & \cellcolor{blue!3}19/7 & \cellcolor{blue!3}14/4 & \cellcolor{blue!3}13/3 & \cellcolor{blue!3}17/6 & \cellcolor{blue!3}0.68 \cr
\cellcolor{blue!8}13. My SIMBA \textcolor{gray}{(15)} & \cellcolor{blue!8}21/6 & \cellcolor{blue!8}17/4 & \cellcolor{blue!8}14/2 & \cellcolor{blue!8}20/4 & \cellcolor{blue!8}0.78 \cr
\cellcolor{blue!3}14. Google Chat \textcolor{gray}{(7)} & \cellcolor{blue!3}8/1 & \cellcolor{blue!3}\textbf{10/0} & \cellcolor{blue!3}7/2 & \cellcolor{blue!3}8/1 & \cellcolor{blue!3}0.88 \cr
\cellcolor{blue!8}15. Webex \textcolor{gray}{(11)}& \cellcolor{blue!8}19/2 & \cellcolor{blue!8}17/3 & \cellcolor{blue!8}12/3 & \cellcolor{blue!8}20/2 & \cellcolor{blue!8}0.85 \cr
\cellcolor{blue!3}16. Yuu SG \textcolor{gray}{(12)} & \cellcolor{blue!3}11/2 & \cellcolor{blue!3}12/2 & \cellcolor{blue!3}10/2 & \cellcolor{blue!3}12/2 & \cellcolor{blue!3}0.82 \cr
\cellcolor{blue!8}17. SAFRA \textcolor{gray}{(8)} & \cellcolor{blue!8}12/2 & \cellcolor{blue!8}14/2 & \cellcolor{blue!8}13/2 & \cellcolor{blue!8}11/2 & \cellcolor{blue!8}0.84 \cr
\cellcolor{blue!3}18. KKBOX \textcolor{gray}{(9)} & \cellcolor{blue!3}\textbf{9/0} & \cellcolor{blue!3}10/2 & \cellcolor{blue!3}11/1 & \cellcolor{blue!3}10/1 & \cellcolor{blue!3}0.91 \cr
\cellcolor{blue!8}19. ShopBack \textcolor{gray}{(11)} & \cellcolor{blue!8}10/2 & \cellcolor{blue!8}11/2 & \cellcolor{blue!8}10/2 & \cellcolor{blue!8}10/2 & \cellcolor{blue!8}0.80 \cr
\cellcolor{blue!3}20. Viu \textcolor{gray}{(12)} & \cellcolor{blue!3}11/3 & \cellcolor{blue!3}7/1 & \cellcolor{blue!3}7/1 & \cellcolor{blue!3}12/3 & \cellcolor{blue!3}0.78 \cr
\cellcolor{green!8}21. eatigo \textcolor{gray}{(13)} & \cellcolor{green!8}18/4 & \cellcolor{green!8}12/2 & \cellcolor{green!8}9/1 & \cellcolor{green!8}17/5 & \cellcolor{green!8}0.75 \cr
\cellcolor{green!3}22. mydlink \textcolor{gray}{(11)} & \cellcolor{green!3}21/4 & \cellcolor{green!3}17/3 & \cellcolor{green!3}15/3 & \cellcolor{green!3}21/4 & \cellcolor{green!3}0.81 \cr
\cellcolor{green!8}23. Synology \textcolor{gray}{(9)} & \cellcolor{green!8}18/1 & \cellcolor{green!8}14/1 & \cellcolor{green!8}14/1 & \cellcolor{green!8}19/1 &  \cellcolor{green!8}0.94 \cr
\cellcolor{green!3}24. EasyBook \textcolor{gray}{(6)} & \cellcolor{green!3}11/3 & \cellcolor{green!3}7/1 & \cellcolor{green!3}8/1 & \cellcolor{green!3}12/3 & \cellcolor{green!3}0.79 \cr
\cellcolor{green!8}25. CNA \textcolor{gray}{(9)} & \cellcolor{green!8}22/3 & \cellcolor{green!8}18/3 & \cellcolor{green!8}12/2 & \cellcolor{green!8}21/4 & \cellcolor{green!8}0.81 \cr
\cellcolor{green!3}26. YouTrip \textcolor{gray}{(11)} & \cellcolor{green!3}27/4 & \cellcolor{green!3}20/3 & \cellcolor{green!3}17/3 & \cellcolor{green!3}27/4 & \cellcolor{green!3}0.87 \cr
\cellcolor{green!8}27. Reddit \textcolor{gray}{(15)} & \cellcolor{green!8}19/2 & \cellcolor{green!8}18/1 & \cellcolor{green!8}13/1 & \cellcolor{green!8}18/2 & \cellcolor{green!8}0.94 \cr
\cellcolor{green!3}28. Our Song \textcolor{gray}{(8)} & \cellcolor{green!3}18/2 & \cellcolor{green!3}15/2 & \cellcolor{green!3}12/2 & \cellcolor{green!3}20/2 & \cellcolor{green!3}0.88 \cr
\cellcolor{green!8}29. Fave \textcolor{gray}{(8)} & \cellcolor{green!8}22/4 & \cellcolor{green!8}20/2 & \cellcolor{green!8}18/3 & \cellcolor{green!8}19/4 & \cellcolor{green!8}0.81 \cr
\cellcolor{green!3}30. DeviantArt \textcolor{gray}{(11)} & \cellcolor{green!3}21/4 & \cellcolor{green!3}19/2 & \cellcolor{green!3}11/1 & \cellcolor{green!3}20/4 & \cellcolor{green!3}0.83 \cr
\hline
\cellcolor{gray!3}Average value & \cellcolor{gray!3}-- & \cellcolor{gray!3}-- & \cellcolor{gray!3}-- & \cellcolor{gray!3}-- & \cellcolor{gray!3}0.812 \cr 
\hline
\end{tabular}
\end{center}
\end{table}
\\
\textbf{\underline{Results and discussion:}} 
Table~\ref{tab: effectiveness} displays the assessment outcomes regarding the effectiveness of AccessFixer.
The first column indicates the names of our collected apps and the number of GUIs intercepted from them.
The second to fifth columns refer to the number of accessibility issues after using AS, PUMA, MATE, and Xbot to detect the GUIs, respectively.
Notably, the accessibility issues mentioned here and in the subsequent experiments only include small size, narrow interval, and low color contrast, and do not involve other issues.
Each cell in Table~\ref{tab: effectiveness} has data in the form of $X/Y$.
Among them, $X$ denotes the number of accessibility issues detected by each tool in the original GUIs, while $Y$ represents the corresponding count of accessibility issues detected by each tool after fixing GUIs with AccessFixer.
The last column shows the average decrease ratio in accessibility issues for each app, and the last row calculates the average results for all apps.
Besides, the gray-highlighted areas in Table~\ref{tab: effectiveness} correspond to the apps that belong to the first group in our ground truth, while the blue-highlighted areas and the green-highlighted areas belong to the second and third groups.
 
\begin{figure}
\centering
\includegraphics[width=7.5cm]{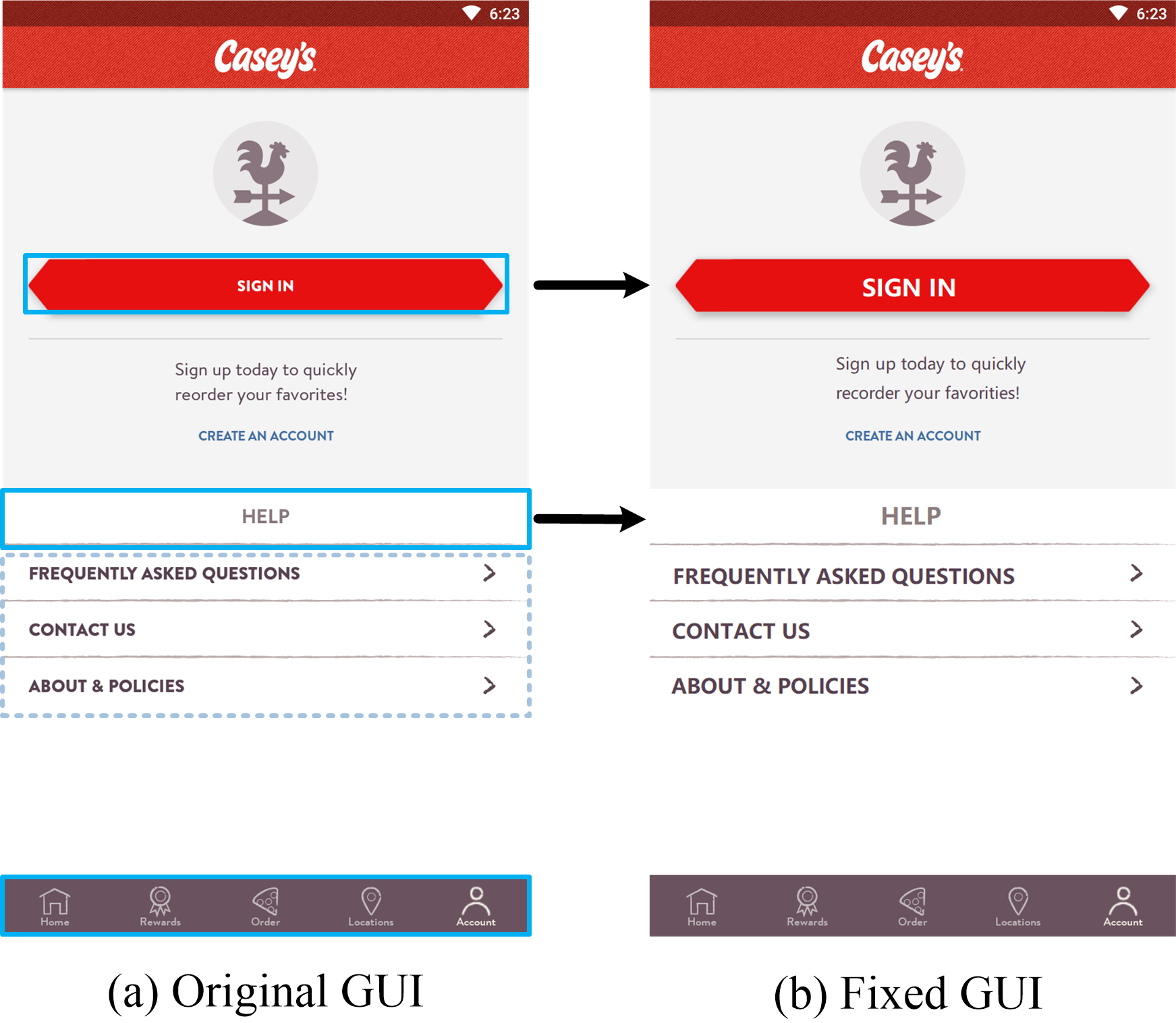}
\caption{The test case of fixed GUIs.}
\label{fig: adjusted}
\end{figure}

The key outcome of our evaluation is that the number of accessibility issues has decreased by 81.2\% on average compared to the original GUIs after AccessFixer fixes these issues. 
Especially for the apps named \emph{KKBOX} and \emph{Google Chat} (bold font in Table~\ref{tab: effectiveness}), their accessibility issues are all solved by our approach after we re-detect them by AS and PUMA.
The primary reason for this lies in that AccessFixer can adaptively adjust the properties of all components in the GUI.
Also, the shapes of all components in these two apps are drawn in a vectorized way, without any components in the form of directly pasted images that our method cannot handle.
For the detection results of 3 app groups, we conduct a \emph{Wilcoxon rank sum test}, and the result shows that there is no significant difference among these groups ($p-value = 0.025, \alpha = 0.05$).
It indicates that AccessFixer can be well applied to apps with different popularity, and make a good performance.
Figure \ref{fig: adjusted} presents a practical comparison case between the original GUI (a) and its fixed result (b).
It can be observed that the size and color contrast of the marked components have been fixed to some extent.
From a visual standpoint, it can be observed that in the fixed GUI, the associated texts below the ``\emph{HELP}'' are also adjusted accordingly to ensure uniform intervals and adequate sizes.

Furthermore, we now analyze the different types of accessibility issues fixed by AccessFixer in our collected ground truth. 
Quantitatively, AccessFixer achieves an average repair rate of 83.75\% for addressing size-related issues, 80.7\% for resolving excessively narrow interval issues, and 80.3\% for addressing color-related issues.
It demonstrates that AccessFixer has a balanced and well-rounded performance in fixing the three kinds of accessibility issues, thereby demonstrating its effectiveness in enhancing overall accessibility for GUIs.
Besides, AccessFixer exhibits a slightly higher repair rate for size-related issues compared to other types.
The reason behind this is that size-related issues are solely associated with the \emph{bounds} attributes of components, and in contrast to issues with narrow intervals and low color contrast, it entails a more direct and quantifiable relationship with the specific attributes.

We then discuss the cases that AccessFixer fails to fix.
One case is shown in the bottom navigation bar in Figure~\ref{fig: adjusted}, the small size issue of component text information in this bar has not been effectively resolved.
The main reason for this is that these views are presented as images rather than \emph{$<$textviews$>$}, making it impossible to directly resize the text within them.
Similarly, AccessFixer is unable to fix components that have missing attributes in the layout-related files, such as the hyperlinks that appear in the source code as \emph{$<$href$>$}, and the predefined global text boxes named \emph{$<$textarea$>$}.
The rationale is that the model cannot predict links for such components, and their signal value would not change in any way.

Concerning these failure cases, one solution is to use image processing methods~\cite{Zhang2016BeyondAG}~\cite{Cruz2018NonlocalityReinforcedCN} to enlarge the content of components.
Yet, this approach may cause components to exceed the boundary and occlude other information.
Another possible strategy involves adding interaction proxies above the problematic components and redrawing their content within those proxies.
Therefore, when users operate the GUIs, they are concerned with the content of proxies, regardless of the accessibility issues in the underlying components.

\subsubsection{The comparison experiment}\label{sub: comparable}
This experiment aims to compare the performance of our proposed AccessFixer and Alotabi et al.'s tool~\cite{Alotaibi2021AutomatedRO} and a human expert’s practical operation.
Alotabi et al.'s tool is the first tool that offers adjustments for GUI components experiencing size-related issues as far as we know.
To conduct this comparison experiment, we re-implement the baseline method, and use it to fix the 197 components with size-related issues in our collected 284 GUIs in Section~\ref{sub: effectiveness1}.
Since the baseline tool can only fix size-related accessibility issues, to ensure a fair experiment, we only consider this kind of issue in the comparison.
Notably, the accessibility issues of narrow interval and low color contrast fixed by AccessFixer are a huge advantage over the baseline tool.
Further, we invite a human expert from the industry with three years of experience in accessibility development to try and fix the accessibility issues in the ground truth we have built. 

We conduct this comparison experiment with the process below.
First, we use the AS to check the GUIs fixed by the baseline tool, and supplement the detection outcomes to the results we have already collected.
Then, following whether size-related issues are successfully fixed and whether the fixed GUIs have new accessibility issues, we classify the detection results into four categories.
The first category (\emph{Fixed}) represents that the components have been successfully repaired by each method, and AS no longer detects any issues in them.
The second category (\emph{Half-baked}) refers to components that have not been fully fixed despite having their attributes adjusted so that the AS can still regard them as inaccessible items.
The third category (\emph{Unfixed}) is the components that our method or baseline tool cannot fix.
The last one (\emph{Extra}) indicates the newly generated problematic components in the fixed GUIs.
Finally, we quantitatively compare the number of these four categories of components in AccessFixer and the baseline tool, together with qualitatively analyzing the visual performance of fixed GUIs.

As for comparing our method with the human expert, it is unrealistic and would take a lot of time and effort for this expert to fix all the 30 apps in the dataset. 
So, we ask him/her to randomly select 3 apps, namely ClassDojo, Webex, and Reddit, to fix based on his/her experience and give him/her 2 weeks to do so. 
After the repair, we adopt four detection tools (AS, PUMA, MATE, and Xbot) to check the repaired GUI and compare the results with AccessFixer.
\begin{figure}
\centering
\includegraphics[width=8.8cm]{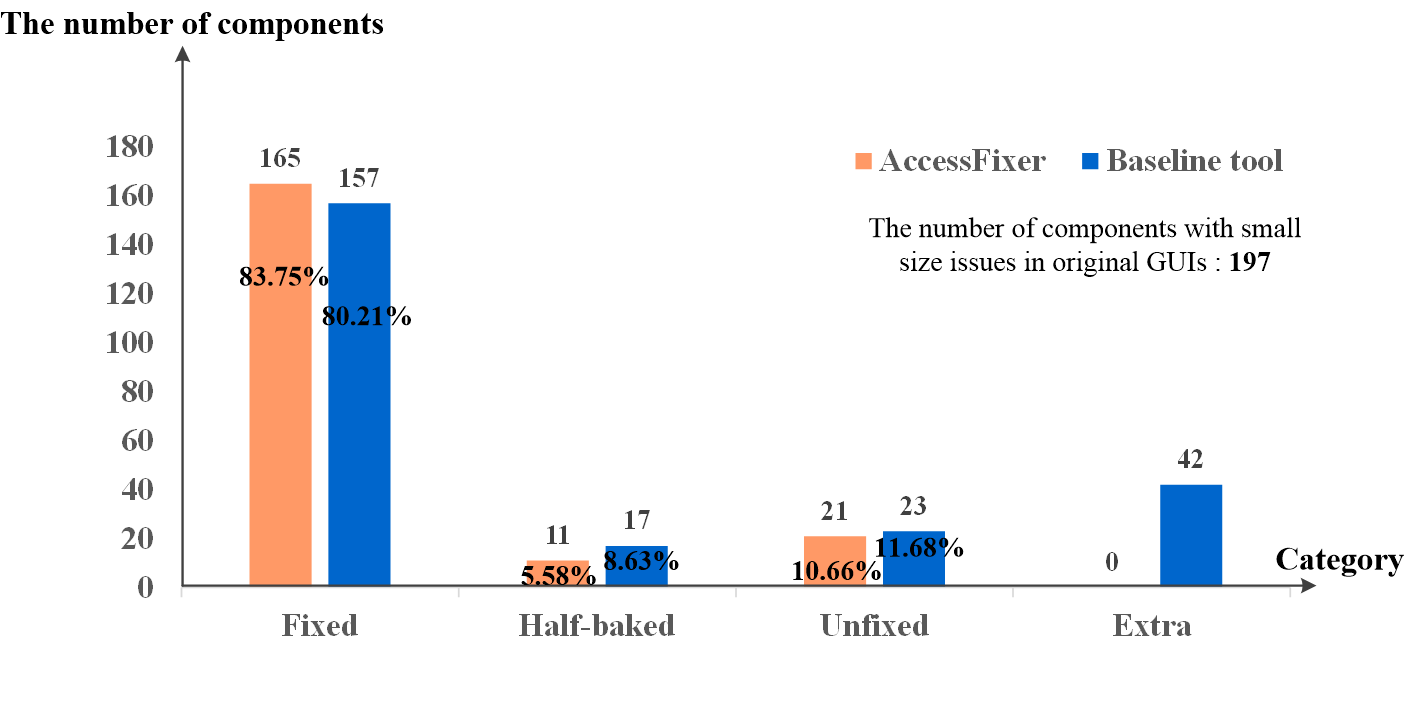}
\caption{Comparison results of AccessFixer and baseline tool.}
\label{fig: compare}
\end{figure}
\\
\textbf{\underline{Results and discussion:}} Figure~\ref{fig: compare} presents the comparison results of AccessFixer and baseline tool.
The abscissa axis represents the four categories (Fixed, Half-baked, Unfixed, and Extra) we defined, and the ordinate axis is the number of components for each category.
The orange bars represent the results of AccessFixer, while the blue bars indicate the results of the baseline tool.
The percentages of \emph{Fixed, Half-baked, }and \emph{Unfixed} in the original 197 problematic components are shown in the middle areas of these bars.
\begin{figure}
\centering
\includegraphics[width=7.5cm]{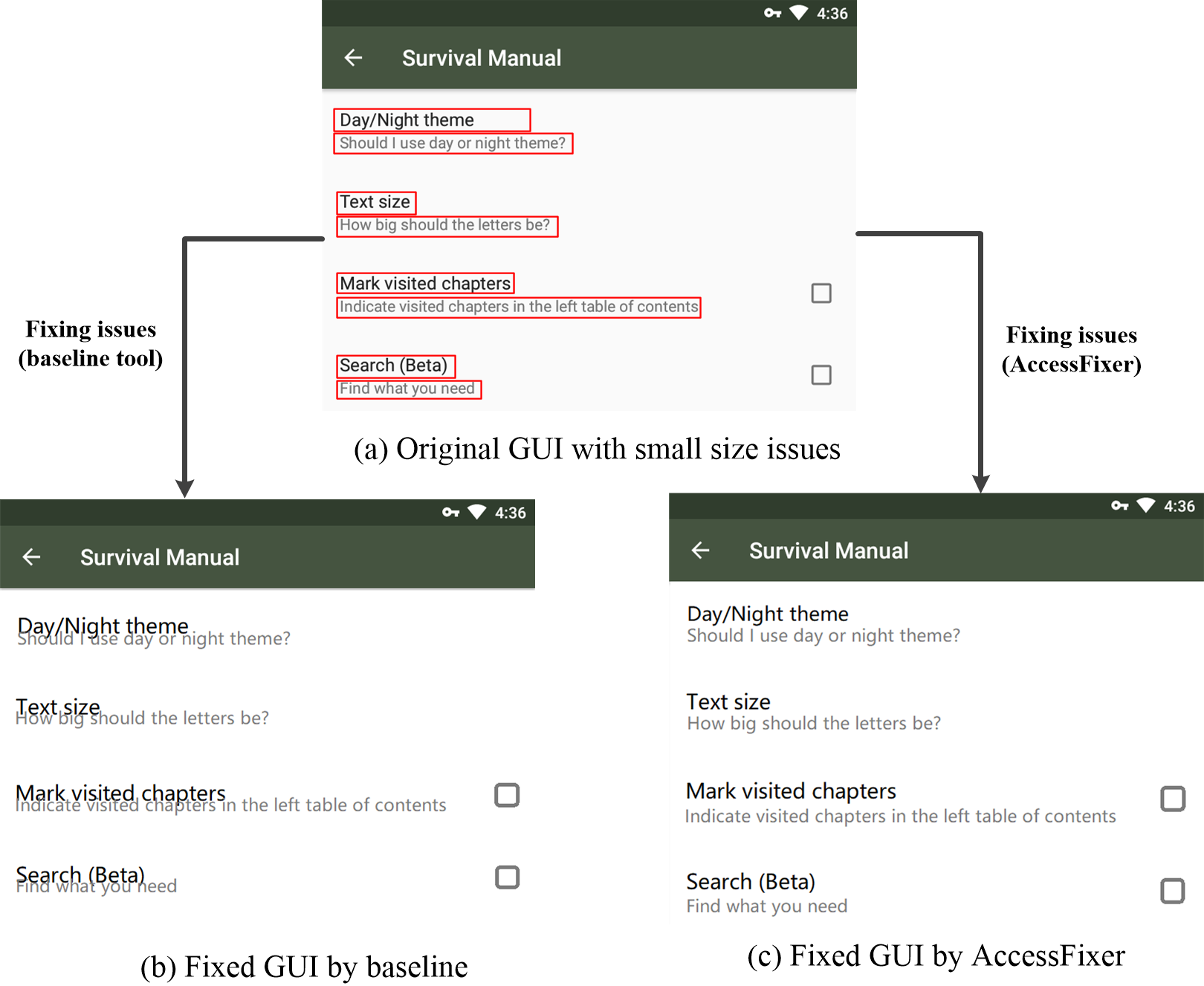}
\caption{The visual performance of baseline and AccessFixer.}
\label{fig: comparevisual}
\end{figure}
As we can see, the AccessFixer is able to fix 3.54\% (83.75\%$-$80.21\%) more components with size-related issues than the baseline tool.
This is not a significant improvement, only an indication that these two methods are comparable.
The primary advantage of AccessFixer versus the baseline tool in fixing size-related issues is that our method would not bring any extra accessibility issues, whereas the baseline tool causes up to 42 extra interval-related issues.
In reality, in the experiment of Section~\ref{sub: effectiveness1}, AccessFixer also does not introduce any new accessibility issues when fixing other types of issues.
Furthermore, we observe the visual performance of both the baseline tool and our method, with an illustrative example shown in Figure~\ref{fig: comparevisual}.
The original GUI (Figure~\ref{fig: comparevisual} (a)) has size-related accessibility issues, marked with red boxes.
Figure~\ref{fig: comparevisual} (b) displays the results after repairing by the baseline tool. 
It is evident that while the size of components has increased, there are instances of compression and overlap among individual components.
In contrast, the GUIs fixed by AccessFixer (Figure~\ref{fig: comparevisual} (c)) maintain uniform intervals without crossing boundaries or hiding other information.

Such performance achieved by AccessFixer benefits from the adaptive mechanism of the R-GCN~\cite{Schlichtkrull2017ModelingRD}, which could collaboratively adjust the attributes of all components without setting the weights.
In the research of fixing GUI accessibility issues, it is important not only to focus on whether a certain type of issue can be successfully fixed, but also to consider how to avoid introducing new issues while achieving successful repairs, together with ensuring a consistent color palette, uniform intervals, and adequate size changes.

AccessFixer and human experts present comparable outcomes in fixing accessibility issues within the GUIs. 
Quantitatively, the human expert achieves repair rates of 85\% for ClassDojo, 89\% for Webex, and 97\% for Reddit, while AccessFixer achieves the rates of 78\%, 85\%, and 94\%, respectively, for these three apps. 
On average, the repair rate by this human expert in these three apps is 4.67\% higher than that of AccessFixer, indicating a slight improvement. 
That is to say, although our method does not achieve the best results as human experts, there is only a small gap between the overall results, which shows that our method remains effective and valuable to some extent.

\subsection{RQ2: Evaluation of the usefulness}\label{sub: usefulness}
Evaluating the usefulness of AccessFixer involves the following two experiments. 
One is to deploy the fixed results to 10 open-source apps in GitHub, recording and analyzing the decisions and comments from developers.
The other is to show the fixed GUIs and original items to 10 low vision users, and then analyze their opinions.

\subsubsection{Experiment with submitting PRs}
In this experiment, we collected the top-ranked apps in F-Droid's~\cite{F-Droid}~\footnote{An installable catalogue of FOSS (Free and Open Source Software) applications for the Android platform.} recommended list from the 10 domains on October 5, 2022.
The rationale for this selection is that these apps attract high attention, so the PR we submitted might be processed in time.
We then install these apps in the Android emulator, fix the GUIs of these apps using AccessFixer, and submit the results by PRs to their GitHub repositories.
Afterward, we would continuously focus on these PRs to record the comments and decisions from developers, engaging in discussions to evaluate the usefulness of our method based on their feedback.
\begin{table}\footnotesize
\renewcommand{\arraystretch}{1.5}
\caption{The decisions of our submitted \emph{PRs}.\label{tab: pull requests}}
\begin{center}
\begin{tabular}{lccc}
\hline
\rowcolor{gray!2}\textbf{apps}  & \textbf{PR\_Id} & \textbf{Status} & \textbf{Repaired Issues} \cr
\hline
	\rowcolor{gray!8}1. Book Reader & \#PR254 & Pending \textcolor{red}{$(\times)$} & Size, Color \cr
	\rowcolor{gray!3}2. Syphon & \#PR272 & Fixing \textcolor{brown}{$(-)$} & Size, Color \cr
	\rowcolor{gray!8}3. NewPipe & \#PR9207 & Merged \textcolor{green}{$(\checkmark)$} & Interval \cr
	\rowcolor{gray!3}4. Timber & \#PR481 & Merged \textcolor{green}{$(\checkmark)$} & Size, Interval \cr
	\rowcolor{gray!8}5. Minimal-Todo & \#PR151 & Fixing \textcolor{brown}{$(-)$} & Color \cr
	\rowcolor{gray!3}6. CoCoin & \#PR60 & Fixing \textcolor{brown}{$(-)$} & Color, Interval \cr 
	\rowcolor{gray!8}7. DroidRec & \#PR56 & Fixing \textcolor{brown}{$(-)$} & Size, Interval \cr
	\rowcolor{gray!3}8. Tuner & \#PR42 & Pending \textcolor{red}{$(\times)$} & Color \cr
	\rowcolor{gray!8}9. AntennaPod & \#PR6424 & Fixing \textcolor{brown}{$(-)$} & Size, Color \cr
	\rowcolor{gray!3}10. SuntimesWidget & \#PR686 & Merged \textcolor{green}{$(\checkmark)$} & Size \cr
\hline
\end{tabular}
\end{center}
\end{table}
\\
\textbf{\underline{Results and discussion:}} Our AccessFixer reports 10 fixed GUIs, and we submit these results by PRs to GitHub repositories and continuously communicate with developers.
A total of 15 developers participated in comments on our submitted PRs.
Table~\ref{tab: pull requests} summarizes the status of these submitted PRs.
Encouragingly, 8 PRs are confirmed by developers, of which 3 confirmed PRs have been merged while the remaining 5 PRs are under fixing. 
The other 2 PRs (\#PR254 in \emph{Book Reader} and \#PR42 in \emph{Tuner}) are still pending, most likely due to the inactive maintenance of the projects.
As follows, we show and discuss three representative comments.
\\
\textbf{C1:} ``\emph{\textcolor{c1}{The GUI you fixed looks very good, we would update it in the next version as much as possible in this form.}}'' \textbf{(Commendatory)}

Among the 15 developers' responses we recorded, 9 of them endorse the fixed results by AccessFixer and express their willingness to merge our submitted PRs or optimize the GUIs in the next version update.
This indicates that AccessFixer can provide developers with a reliable auxiliary way to help them fix accessibility issues, thereby improving GUI accessibility and user experience.
\\ 
\textbf{C2:} ``\emph{\textcolor{c1}{Your suggestion is very good, but we don't know that there are low vision users who will use our app.}}'' \textbf{(Unthoughtful)}

Developers sometimes just approve of our fixed results, but they don't realize that their apps might be used by people with low vision.
As a result, they design the GUIs without considering the accessibility of apps. 
Our proposal also gives hints to developers, guiding them that they should be concerned with the special population and design more friendly GUIs in the subsequent development process.
\\
\textbf{C3:}  ``\emph{\textcolor{c1}{It's appreciated for your suggestions, however, due to the technical restrictions, we have no way to fix all UIs.}}'' \textbf{(Resource-intensive)}

Out of the confirmed PRs, 4 developers expressed their appreciation for us identifying and fixing these issues.
However, these developers deem that implementing such a fix would necessitate significant code refactoring, which would be a time-consuming process and may not yield more benefits.
It's worth noting that all the PRs we submitted were made before the release of ChatGPT 3.5, so developers may have faced the above restrictions when fixing accessibility issues.
Nevertheless, with the assistance of this emerging artificial intelligent technology, utilizing AccessFixer to fix accessibility issues in GUIs can effectively overcome the difficulty of resource-intensive, demonstrating the promising prospects of our approach.

In general, we cannot guarantee that developers of all open-source apps could follow accessibility and incorporate our fixed results, but in the PRs that have been processed, all the developers hold a positive attitude toward our work.
Given these promising comments from developers, we believe that AccessFixer could be a useful approach.

\subsubsection{User study}\label{sub: user study}

Shifting the attention to the user study we conduct, it discusses whether the fixed GUI actually improves the experience for low vision users.
As follows, we present in detail the process of this user study.
\\
\textbf{\underline{Participants recruitment:}} We start by recruiting low vision users to join in our experiment. 
We recruit 10 low vision users (5 males and 5 females; the age ranges from 17 to 33) from a special education college for visually impaired students.
Notably, in contrast to the users recruited for the training set construction (Section~\ref{sub: make}), these participants are newly recruited and only responsible for participating in this user study.
Among them, 8 participants are students in the special education college, and the other 2 participants are teachers.
Every participant receives a \$30 shopping card as a reward after the experiment.
Before the formal experiment, we would conduct a preliminary online communication with these participants, aiming to learn their basic information (i.e., name, age, and occupation), so that we can organize our experiment well.
According to the background survey, all participants have more than 3 years of experience in using apps.
For each recruited participant, we would train them, inform them of our purpose and the process, and carry out this experiment with their consent.
We also promise participants that their basic information will be only utilized in this work rather than anywhere else.
Further, before conducting the user study, we obtain ethical approval from the relevant institutional review board or ethics committee. 
This approval ensures that our study adheres to the ethical guidelines and principles of conducting research involving human participants. 
Also, all participants involved in the user study provided informed consent. 
We provide them with detailed information about the study, its objectives, procedures, potential risks, and benefits. 
Participants are allowed to ask questions and clarify any doubts before giving their consent to participate voluntarily. 
Additionally, we ensure that participants' identities are kept confidential and not shared beyond the research team. 
We also inform the participants about their right to withdraw from the study at any point without any negative consequences.
\\
\textbf{\underline{Experimental setting:}} Our experiment was scheduled for August 19, 2023, and the first, second, and fourth authors of the paper are equally involved in the experiment to fully understand the views of low vision users.
These 10 participants are randomly divided into two groups, and each group has 5 participants.
Then, we evenly divide the original 10 GUIs into two groups, denoted as $g_1$ and $g_2$, and correspondingly, the fixed results of these 10 GUIs are also divided into two groups, labeled as $g_1'$ and $g_2'$.
Regarding the first group of participants, we randomly present them with the GUIs from $g_1$ and $g_2'$.
The other group of participants observes the GUIs in $g_1'$ and $g_2$.
Notably, the participants do not know in advance which is the original and which is the fixed one.
This method can effectively eliminate bias caused by users who may find the problematic components according to their subjective memory.
They then use their weak eyesight to observe these GUIs, and find the components that cannot be seen clearly, and we would record these problematic components.
After participants finish the above process, we first count the number of components with accessibility issues in the original GUIs and the fixed items.
Then, we present them with all GUIs and consult their opinions of our fixed GUIs, and further analyze the recorded insightful responses.
As follows, we demonstrate our comparison and analysis results.
\\
\textbf{\underline{Results and discussion:}} Participants report that the number of components that could be seen clearly in the fixed GUI is higher than that in unfixed GUIs. 
Quantitatively, among the participants in the first group, they identified an average of 7.9 problematic components in the original GUIs, whereas in the repaired GUIs, they found 1.2 problematic components. 
In the second group, they identified an average of 8.4 problematic components in the original GUIs, and in the repaired GUIs, they discovered 1.6 problematic components on average.
These results reveal that GUI fixes with AccessFixer perform better from the perspective of users (average 84.5\% decrease in problematic components).
This also serves as proof of the usefulness of our approach for low vision users.

Meanwhile, we received two valuable and insightful responses from users.
One comment is that ``\emph{\textcolor{c1}{Thank you for designing this tool to improve our experience in using apps, but there seem to be other problems in the GUI also bother me.''}}
Indeed, apart from the three kinds of accessibility issues our method can resolve, there are other types of issues in GUIs that also affect the experience of low vision users.
Yet, these issues may involve different contents, and the solutions for them may also differ.
We are unable to thoroughly solve all the problems within the GUIs in one work, while we try our best to adopt a universal approach to fix three kinds of common problems. 
These statements emphasize the pressing need to repair other types of issues in the GUI, and it is essential to avoid introducing new issues and strive to address various accessibility issues in GUIs simultaneously when proposing new repair methods.

Another one is that ``\emph{\textcolor{c1}{I want other apps on my phone to be able to adapt to the tools you provide. How to implement this?''}}
In fact, both the conventional approaches to detecting and resolving accessibility issues and our novel method primarily center their efforts on app developers, rather than directly catering to the needs of visually impaired users.
Concurrently, the predominant objective of many app vendors is often to maximize profits while minimizing expenses, which can pose significant barriers for marginalized groups in their pursuit of an equitable and optimal user experience.
Hence, as raised by this user's inquiry, a promising approach would be to devise accessibility software that is capable of running natively on Android mobile devices.
First, Accessibility Service~\footnote{An API provided by Android that allows real-time monitoring of GUI display content and supports custom event responses.} provides a way to obtain the displayed content from GUIs in real-time.
Then, several algorithms could be designed to construct optimization schemes for the collected GUI contents.
Afterward, as discussed in Section~\ref{sub: effectiveness1}, regarding the interaction proxies~\cite{Zhang2017InteractionPF} that allow for redrawing the content of components, it could be inserted between the application's original interface and the manifest interface that a person uses to perceive and operate the application.
As such, this method allows third-party developers and researchers to modify the interaction without the application's source code or rooting the phone, all while retaining full system capabilities. 
We believe that such efforts are crucial in enhancing the user experience of low vision users.

\section{Threats to Validity}\label{sec: validity}

One possible risk in our work is that we only collect 30 real-world apps from Google Play to discuss the effectiveness of our method.
Such a small-scale assessment might make it hard to evaluate whether AccessFixer can be applied to various apps.
In reality, we are hard to fully discuss the millions of apps held in Google Play.
Thus, to eliminate this threat, we try our best to select apps with different downloads, popularity, and domains, as reflected in our grouping of apps in Section~\ref{sub: data}.
We deem that such a selection can discuss the effectiveness of AccessFixer while also exploring whether it can be applied to various apps.

Any human involved in the experiments would cause threats.
During the labeling of our dataset for model pre-training, there may be a threat in results due to differences in experience among low vision users. Additionally, our dataset is not entirely labeled by low vision users and is expanded using existing detection tools, which may affect the accuracy of the pre-training model. 
To mitigate this risk, we conduct two rounds of manual observation on selected GUIs without accessibility issues and consult with low vision users to resolve any uncertainties. 
Another potential risk could arise from the overlap between the participants recruited for constructing the training set and those involved in the user study. 
If these sets of users are the same, it might introduce subjective bias into the results of the user survey. 
However, in our two recruitment rounds, we ensure the participants are entirely distinct individuals.
This is done specifically to mitigate any potential impact of such bias on the experimental outcomes.

In our experiment, a potential risk is that we do not compare the differences between the R-GCN model and other deep learning (DL) models in fixing GUI accessibility issues.
In fact, we attempted to use other models, such as Convolutional Neural Network (CNN)~\cite{LeCun1998GradientbasedLA}, Recurrent Neural Network (RNN)~\cite{Liu2021SelfishSR}, GCN~\cite{Kipf2017SemiSupervisedCW}, Graph Attention Network (GAT)~\cite{Velickovic2017GraphAN}, and Graph Auto-Encoder (GAE)~\cite{Kollias2022DirectedGA}, to fix GUI accessibility issues, but all failed. 
The primary reason is that the downstream tasks that the models could perform focus on classification and label prediction.
Therefore, we could not obtain the features to adjust the attributes of components, preventing us from fixing accessibility issues.

\section{Conclusion}\label{sec: conclusion}
The accessibility of GUIs plays a crucial role in determining the accuracy and comfort with which visually impaired users can interact with apps.
Accessibility issues in GUIs seriously hinder the normal operation of these users.
To address this concern, this paper proposes AccessFixer to fix the accessibility issues in terms of small size, narrow interval, and low color contrast.
AccessFixer can formally represent GUIs as graphs, and follow the R-GCN pre-trained model to automatically adjust the attributes of components, thereby achieving the goal of fixing accessibility issues.

Our experiments evaluate the effectiveness and the usefulness of AccessFixer.
Concerning effectiveness, we observe an average 81.2\% reduction in accessibility issues across the 30 real-world apps fixed by AccessFixer.
Also, our method outperforms the baseline tool in terms of the category of issues they could fix, the fixed results, and the visual consistency.
To evaluate the usefulness of AccessFixer, we submit the fixed GUIs with PRs to 10 open-source apps in GitHub.
The results show that 8 of them are well processed, and they have been merged or under fixing.
Also, the results of a user study demonstrate that the AccessFixer is useful for low vision users.

\section{Acknowledgment}
We extend our gratitude to the low vision users who generously offered their time and effort in assisting us with our experiments. 
Their assistance is invaluable in the successful completion of this study.
This work is funded by Jilin Provincial Natural Science Foundation, 20230101070JC, Science and Technology Research Project of Education Department of JiLin Province of China (JJKH20211104KJ), ``the National Natural Science Foundation of China (NSFC) No. 62102160'', and supported by ``the Fundamental Research Funds for the Central Universities, JLU, (2022-JCXK-16)''.

\bibliographystyle{IEEEtran}
\bibliography{accessibility}

\end{document}